\documentclass[aps,pre,twocolumn,superscriptaddress]{revtex4-2}

\usepackage{graphicx}
\usepackage{epstopdf}
\usepackage{natbib}
\usepackage{xcolor}
\usepackage{bm}		

\newcommand{\sciexp}[2]{{#1}\ensuremath{\,\times\,10^{#2}}}

\newcommand{\DHe}{D$^3$He}

\newcommand{\del}[1]{}

\newcommand{\PPPL}{Princeton Plasma Physics Laboratory, Princeton, NJ 08543, USA}
\newcommand{\PAS}{Department of Astrophysical Sciences, Princeton University, Princeton, NJ 08544, USA}
\newcommand{\UMich}{Center for Ultrafast Optical Science, University of Michigan, Ann Arbor, MI 48109}

\begin{document}

\begin{abstract}

Proton deflectometry is increasingly used in magnetized high-energy-density plasmas to 
observe electromagnetic fields.
We describe a reconstruction algorithm to recover
the electromagnetic fields from proton fluence data in 1-D.  
 The algorithm is verified against analytic solutions and applied to example data.
Secondly, we study the role of source fluence uncertainty for 1-D reconstructions.  We show that
reconstruction boundary conditions can be used to constrain the source fluence profile, and 
use this to develop a reconstruction using a specified pair of boundary conditions on the magnetic field.
From these considerations we experimentally demonstrate a hybrid mesh-fluence reconstruction technique where 
fields are reconstructed from fluence data in an interior region
with boundary conditions supplied by direct mesh measurements at the boundary.
\end{abstract}

\title{Proton deflectometry analysis in magnetized plasmas: \\ magnetic field reconstruction in one dimension}
\author{W. Fox}
\thanks{These authors made equal contributions to this work.}
\affiliation{\PPPL{}}
\affiliation{\PAS{}}
\author{G. Fiksel}
\thanks{These authors made equal contributions to this work.}
\affiliation{\UMich{}}
\author{D.B. Schaeffer}
\email{Present address: Department of Physics and Astronomy, University of California, Los Angeles}
\affiliation{\PAS{}}
\author{J. Griff-McMahon}
\affiliation{\PAS{}}
\date{\today}

\maketitle

\section{Introduction}

Proton deflectometry (or radiography) \cite{KuglandRSI2012, BottJPP2017, SchaefferRMP2023} is increasingly 
used to observe the
evolution of electric and magnetic fields in high-energy-density plasmas.
This has enabled magnetic field
observations in
experiments ranging from
compressed fields for inertial 
fusion energy \cite{GotchevPRL2009} and 
self-generated magnetic fields in laser-solid interaction \cite{LiPRL2006, PetrassoPRL2009, GaoPRL2015}, 
to laboratory astrophysics measurements of
Weibel instability \cite{FoxPRL2013}, 
magnetic reconnection \cite{NilsonPRL2006,LiPRL2007b, FikselPRL2014, RosenbergPRL2015}, magnetized shocks \cite{SchaefferPRL2019},
and plasma dynamos \cite{TzeferacosNatComm2018}.

The principle of the measurement,
which has been discussed in Refs.~\cite{KuglandRSI2012, BottJPP2017},
and a recent review article~\cite{SchaefferRMP2023},
is to use a beam of protons to
map the electromagnetic fields in an experiment.
A point source of protons is generated,
through either a laser-driven
implosion of a \DHe{}-filled capsule, or laser-solid interaction.
The protons then stream through a region
under study, where they pick up small-angle deflections
from the electromagnetic fields, before
propagating ballistically to a detector.
The goal is to use the detected protons to infer the electromagnetic fields.
Often times a grid or mesh is used to
break the protons into beamlets (e.g. Refs. \cite{LiPRL2007b, PetrassoPRL2009}).  This has the
virtue that the beamlets can
be directly located on the detector to measure the 
proton final positions. 
X-rays are also generated in \DHe{}
implosions and these can be mapped
simultaneously using appropriately designed
detector stacks \cite{JohnsonRSI2022, MalkoApplOptics2022}.  Since x-rays are 
not deflected by electromagnetic fields, the
x-ray beamlets provide a direct reference of the 
undeflected beamlet locations.

While the mesh enables a direct measurement
of the proton deflections, it also sacrifices
spatial resolution.
To observe at higher resolution, it
is also possible to take direct
proton fluence images without a mesh.  In this
case, the proton focusing and defocusing by the electromagnetic
fields leads to fluence variations on the detector,
and the goal is then to reconstruct the fields which create these variations.
Generating forward proton models 
(e.g. Ref.~\cite{FikselPRL2014}) to compare with 
experimental data is straightforward, since one simply
has to generate model fields and then calculate and
bin proton trajectories to generate an image for comparison.
A quantitative analytic theory connecting proton deflections to
fluence variations was described by 
Kugland \textit{et al} \cite{KuglandRSI2012}.
Finally, and potentially most powerfully, algorithms have
been developed to
invert measured fluence images
to obtain the experimental electromagnetic fields \cite{BottJPP2017, KasimPRE2017, DaviesHEDP2023}.
These inversion techniques generally involve an optimization 
or relaxation-type solution to the Monge-Ampere transport equation \cite{BottJPP2017,KasimPRE2017, SulmanANM2011}, 
while other algorithms have been developed that 
exploit the analogy between deflectometry and charged particle motion \cite{DaviesHEDP2023},
or use neural networks \cite{ChenPRE2017}.

In this paper we develop and verify a 1-D inversion algorithm
to obtain 1-D field profiles from proton fluence profiles 
through direct integration of an ordinary differential equation (ODE).
This is complementary to 2-D algorithms mentioned 
above \cite{BottJPP2017, KasimPRE2017, DaviesHEDP2023}.
The algorithm is fully non-linear and can reconstruct for 
large proton deflections, as long as there are no caustics and the 
proton trajectories do not cross.  It therefore
works in the same ``non-linear injective'' regime of Ref.~\cite{BottJPP2017}.
A virtue of a 1-D algorithm is that it can run
very quickly (typically $<0.1$ sec for a reconstruction), and therefore can be
easily embedded within higher-level workflows for error analysis and parametric scans.
Furthermore, several experiments, including magnetic reconnection \cite{RosenbergPRL2015},
magnetized cylindrical implosions \cite{GotchevPRL2009},
or transport in magnetized plasmas \cite{WalshPoP2020} related to 
fusion concepts such as MagLIF \cite{GomezPRL2020}
have 1-D or nearly-1-D regions
which can be analyzed by this technique.
Variations on this algorithm have been
implemented in cylindrical geometry to analyze 
cylindrically-symmetric expanding plasmas \cite{CampbellPRL2020}.

Secondly, we investigate some subtleties for 1-D reconstructions, such as the connection
between the reconstruction boundary conditions
and the ``source'' proton fluence, which is the proton fluence before it is deflected by the 
experimental fields.
We show that reconstruction boundary conditions
can be used to directly constrain the average source proton fluence.
We use this result to implement arbitrary magnetic field boundary conditions
for the 1-D algorithm.  (Other commonly-used solvers, e.g.
PROBLEM \cite{PROBLEM_2024} at the time of this writing, apply the
zero-deflection boundary condition $\mathbf{B_\mathrm{tangential}} = 0$.)
We also explore the ``integration error'' introduced by
error in the source fluence.  We focus on error in the average level, 
however the results also have implications for the more 
general question of source fluence non-uniformity.
These results complement prior statistical analysis of 
source fluence uncertainties \cite{KasimPRE2019}.
Finally, we 
develop and experimentally demonstrate a ``hybrid'' proton fluence deflectometry
technique, with separate mesh
and fluence regions in the same radiograph, where mesh regions provide 
direct measurement of magnetic field
boundary conditions for the 
reconstruction in the fluence region.

This paper is intended as the beginning of a series of papers
which discuss analysis of proton data for recent
experiments.  The focus is on proton deflectometry
with careful analysis of the various sources of
measurement uncertainty, to allow  
quantitative statements about measured magnetic fields with error bars.
A first physics analysis of magnetic reconnection
experiments using this analysis technique is presented in Ref.~\cite{Fox2023}.
Finally, an appendix describes the implementation
of the routines (presently in Matlab) 
in a package called PRADICAMENT.

\section{Proton deflections}
\label{Section_proton_deflections}

\begin{figure}
\centering
\includegraphics{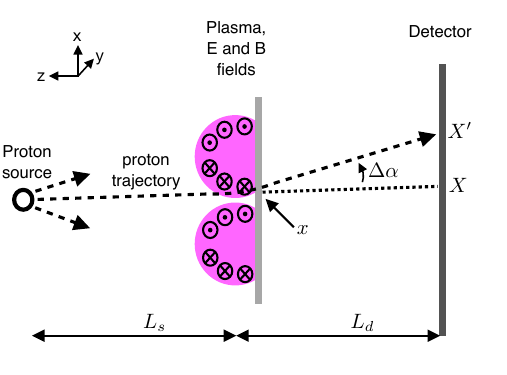}
\caption{Schematic of the measurement setup and coordinate system.  A proton 
source produces a point-source of high-energy protons.   These stream
through an experimental region, where the protons 
pick up deflections $\Delta \alpha$ due to electromagnetic fields,
after which they propagate ballistically to a detector.  
An example proton crosses the plasma plane at position $x$, where 
the electromagnetic fields deflect the trajectory, causing the 
proton to arrive at position $X'$ rather than $X$ on the detector.}
\label{Fig_Setup}
\end{figure}

In this section we briefly review the measurement
setup and basic theory of proton deflections to fix the geometry and notation to be
used below.
The reader is referred to Refs.~\cite{KuglandRSI2012, BottJPP2017, SchaefferRMP2023}
for extended discussion of the proton deflectometry theory.
Figure~\ref{Fig_Setup} shows a typical
experimental geometry.  
The protons emerge from a 
point-source located at a distance $L_s$ from the plasma under study.  The detector is 
positioned at a distance $L_d$ on the opposite side.
(The figure is not to scale, as often $L_d \gg L_s$, in a point-magnification geometry.)
As the protons travel through the plasma region, they pick up
small angle deflections $\Delta \alpha$ due to 
electromagnetic fields.   The figure shows
an example magnetic geometry representative of 
recent experiments on laser-driven magnetic reconnection \cite{LiPRL2007b, RosenbergPRL2015},
in connection with the verification example below.

Following the development from prior
work on proton deflectometry \cite{KuglandRSI2012, BottJPP2017},
we consider high energy protons which only pick up small deflections
$|\Delta\mathbf{V}| \ll V_p$ while propagating through the plasma,
where $V_p = (2 E_p / m_p)^{1/2}$ is the proton speed given
the initial proton energy $E_p$.  In this limit, the deflection 
is given by an integral over the straight-line trajectory,
\begin{equation}
\Delta \bm{\alpha} = \frac{\Delta \mathbf{V}}{V_p}  = \frac{e}{m_p V_p^2} \int \mathbf{(E + V_p \times B)}_{\perp} \, d \bm{\ell}.
\end{equation}
The validity of this limit is extensively discussed
in Refs.~\cite{KuglandRSI2012} and \cite{BottJPP2017}.

Considering the detector geometry,
the proton crossing the plasma at position $x$ arrives at the detector at position
\begin{equation}
\mathbf{X'} = (1+L_d/L_s) \mathbf{x} + L_d \Delta \bm{\alpha}(\mathbf{x}).
\end{equation}
It is convenient to work just in the coordinate system of the plasma plane,
so we use the ``final'' proton position mapped back to the plasma plane,
$\mathbf{x'}=\mathbf{X'}/M$, using the magnification $M = (1+L_d/L_s)$,
so that 
\begin{eqnarray}
\mathbf{x'} & = & \mathbf{x} + \frac{L_s L_d}{L_s + L_d} \Delta \bm{\alpha} \\
 & = & \mathbf{x} + \bm{\xi}(\mathbf{x}),
\label{Eq_mapping_2d}
\end{eqnarray}
introducing the deflection $\bm{\xi}(\mathbf{x})$.
With this definition, 
\begin{equation}
\bm{\xi}(\mathbf{x}) = K_B^{-1} \int d \bm{\ell} \times \mathbf{B} + K_E^{-1} \int \mathbf{E}_\perp \,d\ell,
\end{equation}
using the deflection ``rigidity'' factors
\begin{equation}
K_B = \frac{m_p V_p}{e} \frac{L_s + L_d}{L_s L_d},
\label{Eq_KB}
\end{equation}
and
\begin{equation}
K_E = \frac{m_p V_p^2}{e} \frac{L_s + L_d}{L_s L_d}.
\label{Eq_KE}
\end{equation}
The units of $K_B$ and $K_E$ are conveniently (in SI) Tesla and V/m.
However, with this formulation, the mapping can work in any consistent unit scheme, 
and for example, $K_B$ can be converted to Gauss. 
If $\int B_y\, dz$ is given in T-m and $K_B$ in T, this produces a deflection $\xi$ with units of m.
The interpretation of $K_B$ is that, for example, given $K_B = 50$~T, for a line integrated field $\int d\ell \times B$ = 50~T-mm,
the proton will be deflected 1~mm in plasma plane units.

For the present analysis, we now specialize to a one-dimensional
geometry, with protons propagating primarily along
$z$, and deflected \textit{only} in the $x$-direction, so that we
have a mapping which is the 1-D version of Eq.~\ref{Eq_mapping_2d},
\begin{equation}
x' =  x + \xi(x),
\label{Eq_mapping}
\end{equation}
where $\xi$ is a function of $x$ only, given by
\begin{equation}
\xi(x) = K_B^{-1} \int B_y\,dz + K_E^{-1} \int E_x\,dz. 
\end{equation}

The proton fluence (defined as protons / unit area, or a similar quantity) 
is assumed first to have a known 
``initial'', or ``source'' fluence $I_0(x)$,
when the protons first reach the plasma plane.  The proton deflections $x \to x'$
then maps this fluence to the detector image $I(x')$.
For a given magnetic field structure, one can calculate
 a synthetic proton fluence image, which we call $I_{fwd}$.
 To do this, one calculates many proton mappings via Eq.~\ref{Eq_mapping}, with initial positions
drawn from the source fluence profile $I_0(x)$, binning the 
final positions $x'$ to determine the final fluence profile.

In 1-D, the fluence transforms according to the Jacobian
of the proton mapping \cite{KuglandRSI2012},
\begin{equation}
I(x') = \frac{I_{0}(x)} {| dx'/dx |}.
\label{Eq_fluence}
\end{equation}
This equation holds if the magnetic field is limited in magnitude to an extent that
 $dx'/dx > 0$, which is equivalent to the absence of
caustics in the proton image, i.e. that proton trajectories do not cross \textit{en route} to the detector.
While Eq.~\ref{Eq_fluence} is valid in non-caustic regimes, the forward binning technique
works even in caustic regimes and is therefore more general.
Ref.~\cite{KuglandRSI2012} discusses caustic formation 
extensively and provides several examples.

Equation~\ref{Eq_fluence}
is equivalent to a statement of conservation of protons,
\begin{equation}
\int_{x_1}^{x_2} I_0(x) \,dx = \int_{x'_1}^{x'_2} I(x') \,dx'
\end{equation}
considering an integral on $[x_1,x_2]$ of the initial protons
or $[x'_1,x'_2]$ over the final protons.
A perennial subtlety of the 
analysis is that the fluence data $I(x')$ is ``observed at'' the final coordinates $x'$,
but depends on the source fluence at the initial coordinates, $I_0(x)$.

 \section{Reconstruction}

We now develop how to reconstruct the magnetic (or alternatively electric) fields from the fluence data.
The equations are the 1-D limit of prior image-fluence relations \cite{KuglandRSI2012, BottJPP2017}, however with the
1-D formulation, the present method departs from the relaxation method of Refs. \cite{BottJPP2017, SulmanANM2011}.
We introduce $b(x) = \int B_y \,dz$ as the line-integrated magnetic field, for brevity.
Hereafter, we also assume the electric deflection is negligible, so that there is a
constant relation between $\xi$ and $b$.    For the more general case, one can analyze
multiple reconstructions from various orientations \cite{PetrassoPRL2009} 
or with different proton energies to separate the electric and magnetic field contributions.

To reconstruct, we determine the relationship between the mapping and the measured proton fluence $I(x')$.
First, we find, using Eq.~\ref{Eq_mapping}, 
\begin{equation}
\frac{dx'}{dx} = 1 + \frac{d\xi}{dx}.
\end{equation}
After substituting this in Eq.~\ref{Eq_fluence}, and using $b = K_B \xi$, we find the following relation between
$b$, $I$ and $I_0$,
\begin{equation}
\frac{db}{dx} = K_B \frac{d\xi}{dx} = K_B  \left( \frac{I_0(x)} {I(x')} - 1 \right).
\label{Eq_Bderiv}
\end{equation}
The simple form of Eq.~\ref{Eq_Bderiv} is rather deceptive since the \textit{LHS} of the equation has the 
magnetic field as a function of the initial proton coordinates $x$ while the \textit{RHS} depends
as well on the final proton coordinates $x'$ through $I(x')$, which are in turn coupled through Eq.~\ref{Eq_mapping}.
This being said, we have now obtained a differential equation, Eq.~\ref{Eq_Bderiv}, 
relating the line-integrated $B$ field to the observed proton fluence.
For the solution, we regard $I(x')$ and $I_0(x)$ as input data, and integrate to obtain $b(x)$.
Numerical solutions are straightforward using ODE solvers, either by hand or using pre-built packages.

Since Eq.~\ref{Eq_fluence} relies on the assumption that deflections are
sufficiently weak that $dx'/dx > 0$,
the present reconstruction algorithm requires the same condition.  
This regime was called the ``non-linear injective regime'' in Ref.~\cite{BottJPP2017}.
This regime guarantees that for each proton final position $x'$ the protons arrived from only one $x$.
(Ref.~\cite{BottJPP2017} also describes a yet-weaker-deflection regime called the ``linear" regime.  
The present reconstruction technique is also valid in the linear regime.)

Given the structure of Eq.~\ref{Eq_Bderiv}, a \textit{unique} reconstruction will be
provided by the solution to the equation plus a boundary condition $b_0 = b(x_0)$ at
some specified point $x_0$.
Without the boundary condition, the solution is unique up to 
a uniform offset magnetic field $\bar{b}$ added to the overall solution,
which also results in an additional, uniform spatial offset between $x$ and $x'$.
Equivalently, without a specified boundary condition, 
the reconstruction provides the \textit{relative} change to the magnetic field on the domain.
We return to further discussion of boundary conditions below.

We can also calculate the line-integrated current,
 \begin{equation}
 \int J_z \,dz = \frac{1}{\mu_0} \frac{d}{dx} \int B_y \,dz = \frac{K_B}{\mu_0} \left( \frac{I_0(x)} {I(x')} - 1 \right).
 \label{Eq_Jparallel}
 \end{equation}
 This shows the direct relationship between the parallel current density and the 
 proton fluence.    The subtlety here again is the mapping $x \to x'$, so that 
 for a current density at $x$, the associated proton fluence is observed at $x'$.
 This relation is the 1-D analog of equations derived in Ref.~\cite{GrazianiRSI2017}. 
Equation~\ref{Eq_Jparallel} shows the close 
 relationship between the proton fluence and current density, and therefore
 why an additional integration is needed to obtain the magnetic field, 
 via Eq.~\ref{Eq_Bderiv}.

\section{Verification}

\begin{figure}
\centering
\includegraphics{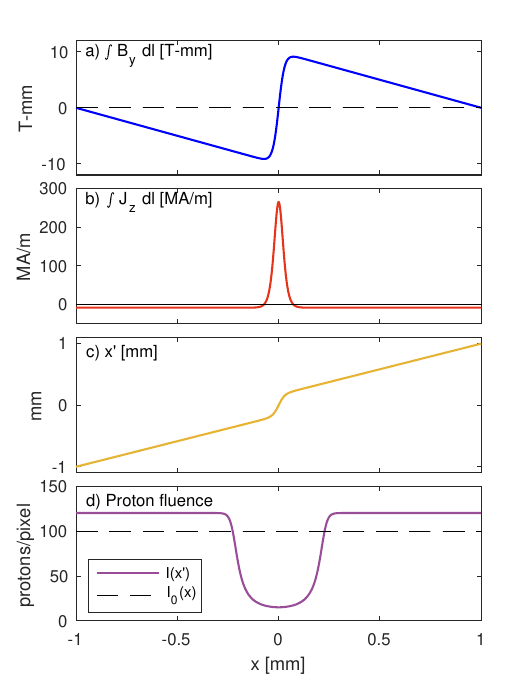}
\caption{Analytic profiles for reconstruction verification.   (a)~Line-integrated magnetic field
profile.  (b)~Plasma current.   (c) Proton final position $x'$ for each $x$.  
(d) Proton fluence profile}
\label{Fig_Analytic}
\end{figure}

We next demonstrate a verification of the reconstruction technique using an example analytic set of 
fields.  We choose a magnetic field profile which is representative of
current sheet formation between colliding magnetized plasmas in magnetic reconnection experiments \cite{FoxPRL2011}.
From  the analytic field profiles, we  calculate a synthetic proton fluence.  This proton
fluence is then fed (numerically) into the reconstruction algorithm, and we verify that the 
reconstruction matches the analytic magnetic field.

\begin{figure}
\centering
\includegraphics{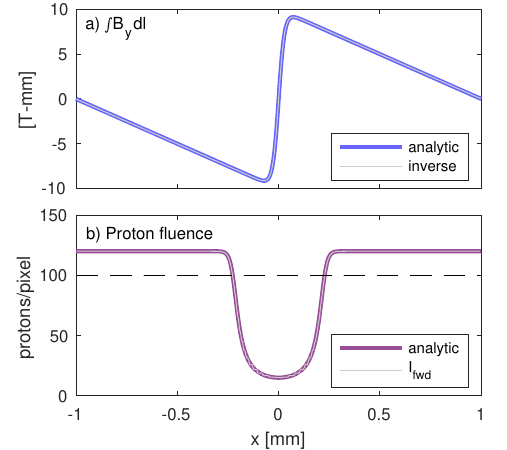}
\caption{Reconstruction demonstration.  (a) Magnetic field profiles comparing the reconstructed
magnetic field with the original analytic form.  (b)  Proton fluence profiles, comparing the 
given input proton fluence, and a forward
proton model from the reconstructed magnetic field profile $I_{fwd}$.}
\label{Fig_Reconstruction}
\end{figure}

We assume a magnetic profile of the form
\begin{equation}
b(x) =  b_0 \tanh (x/\delta) \; \left(1 - \frac{|x|}{L_B} \right).
\end{equation}
Here $b_0$ is the peak (line-integrated) magnetic field which we take as 10~T-mm.
$L_B$ is a constant related to the distance to 
center of each magnetized bubble, which we take as 1~mm,
so that the B field returns to zero at $x= \pm L_B$.   We will
use this formula, and reconstruct, over the domain $x \in [-L_B, L_B]$.  Finally the 
current sheet width parameter $\delta$ is taken as 30~$\mu$m.
This analytic profile is shown in Fig.~\ref{Fig_Analytic}(a).
From this, we calculate a line-integrated plasma current density $\int J_z\, dz = (1/\mu_0)\, d/dx \int B_y\,  dz$,
shown in Fig.~\ref{Fig_Analytic}(b).  The current shows a strong
positive spike in the current sheet with a magnitude larger than 250~MA/m.
Away from the current sheet, the current density is slightly negative, representing a return current.

We next produce a synthetic set of 1-D proton fluence data, using typical proton parameters
from experiments.   We take $E_p = 14.7$~MeV,  $L_s = 10$~mm,
and $L_d = 150$~mm, from which we evaluate $K_B = 59$~T\@.
From this, we calculate proton deflections $\xi$ and accordingly $x'$ as a function of $x$,
which is shown in Fig.~\ref{Fig_Analytic}(c).   Finally, we calculate the
proton fluence, based on a nominal uniform source proton fluence of  $I_0$ of 100 protons/pixel,
shown in Fig.~\ref{Fig_Analytic}(d).
The final proton fluence can be calculated semi-analytically using the
mapping $x \to x'$ and analytic $d\xi/dx$, or numerically 
by binning the final proton positions.
The reversal of the magnetic field causes the protons on
opposite sides of the current sheet to diverge (as illustrated in Fig.~\ref{Fig_Setup}),
producing a broad proton fluence depletion near $x=0$.

This proton fluence profile is then used in the inversion procedure described above,
which is to numerically integrate Eq.~\ref{Eq_Bderiv} coupled to Eq.~\ref{Eq_mapping}.
We use the synthetic proton fluence $I(x')$ shown in Fig.~\ref{Fig_Analytic}(d)
as input data.
Secondly, at this point we assume we know the source proton fluence $I_0$ = 100 protons/pixel,
and the initial condition on the magnetic field $b=0$ at $x=-1$~mm, and
apply these in the reconstruction.
(The reconstruction can equally start at other initial conditions, such as $b=0$ at $x=+1$~mm;
we verified the inversion produces identical reconstructions within the tolerances in each case.)

The results are shown in Fig.~\ref{Fig_Reconstruction}(a), showing excellent agreement between
the analytic profile and the reconstruction.
The maximum deviation between the inversion and the analytic $B$ profile is $<$\sciexp{1}{-5}~T-mm,
or less than 0.1\%.

Finally, as a standard check, we calculate the forward proton fluence $I_{fwd}$ from
the reconstruction.  
The comparison of $I_{fwd}$ with the input data $I(x')$
is a useful (and minimal) test for the general case 
when there is no analytic magnetic field to compare against.
To do so, we numerically calculate $\xi$ based on the reconstructed field 
and bin the final proton positions, shown
in Fig.~\ref{Fig_Reconstruction}(b).
The agreement with the original input $I(x')$ is excellent, which is 
to be expected since we also had agreement between the analytical and reconstructed fields.
Some fine-scaled ``jaggedness'' can be observed in $I_{fwd}$, which is 
due to the finite spatial resolution of the reconstruction.  The maximum difference
between $I_{fwd}$ and the synthetic input data was $<2.5$~protons/pixel, and
 the RMS deviation was $<0.5$~protons/pixel.
These results illustrate the overall numerical verification of this 1-D reconstruction technique.

\section{Boundary conditions and the source fluence $I_0$}

We now study how the source proton fluence and boundary conditions can impact 1-D magnetic reconstructions.
In this section we develop the relationship between these
quantities and how uncertainties in these quantities
feed through to results of the reconstruction.

\begin{figure}
\centering
\includegraphics{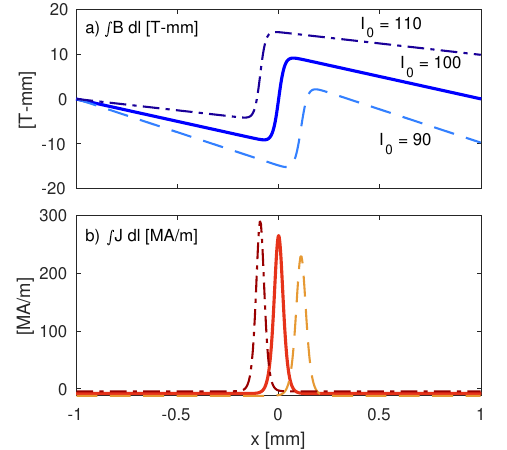}
\caption{Proton reconstructions under multiple values of $I_0$.   a) Magnetic field
calculations for uniform $I_0 = $(90, 100, and 110 protons/pixel as labeled.
b) Associated plasma current density reconstructions.}
\label{Fig_Multi_I0}
\end{figure}

For the reconstruction
above (Fig.~\ref{Fig_Reconstruction}), we assumed we knew two important
quantities: the initial condition to start the integration from,
and secondly the source proton fluence $I_0(x)$.
Solutions of Eq.~\ref{Eq_Bderiv} can add a uniform magnetic field,
which leads to a constant additional offset of $x$ and $x'$,
So, without a specified boundary condition, only the relative change of the 
field across the integration domain is obtained from the analysis.   For some applications, such as observing
the RMS or fluctuating components of the magnetic field (e.g. \cite{BottJPP2017,TzeferacosNatComm2018})
the relative variations may be sufficient.  However other
contexts, such as for magnetic reconnection or collisionless shock experiments, 
the absolute magnitude may be very important.

Secondly, the source fluence $I_0$ is also required 
for the reconstruction.
In the example above, we posited that we knew the source 
fluence $I_0(x) = 100$~protons/pixel.   However, we 
now conduct an exercise imagining there was
some uncertainty in determining this quantity.
Figure~\ref{Fig_Multi_I0} shows the 
results for reconstructing the same $I(x')$ above, however
with $I_0$ now set at 90 or 110 protons/pixel (still uniform on the domain).
The resulting reconstructed fields are shown in Fig.~\ref{Fig_Multi_I0}(a), as the 
dashed and dot-dashed curves, labeled with the associated $I_0$.
We observe that this change in $I_0$ leads to significantly different reconstructed magnetic fields.
Roughly speaking, the change in $I_0$ introduces a positive
or negative ramp to $b(x)$, which we will call the ``integration error.''
The magnetic field still makes a similar jump in each case,
however the position of the jump is offset, due to the error propagated through the 
mapping $x \to x'$.
By the right-hand boundary, the
curves have significantly different magnitudes, yielding $b(x)$
near $\pm 10$~T-mm,
which is approximately equal to the \textit{maximum} value $b_{max}$
from the $I_0=100$ solution.  It is clear that $|\delta b| / |b_{max}| \sim~$100\%,
where $\delta b$ is the difference between solutions.

Secondly, Fig.~\ref{Fig_Multi_I0}(b) shows the 
associated reconstructed current density.  We see 
that the location of the peak currents are offset spatially, just like the 
spatial offsets of the magnetic field jumps.
However, on the other hand, the \textit{peak} currents are actually fairly similar 
($\pm$10\%).  This indicates that some outputs from an
analysis, such as peak current, can be more robust to $I_0$
errors than magnetic field measurements.

We now characterize the ramp (``integration error'') resulting from an error in $I_0$. 
We return to Eq.~\ref{Eq_Bderiv},  and introduce an error in the 
source fluence measurement $\delta I_0$.  We obtain
\begin{equation}
\frac{db}{dx} = K_B \left( \frac{I_0}{I(x')} + \frac{\delta I_0}{I(x')} - 1 \right).
\end{equation}
While this equation is complicated, owing again to the 
non-linear dependence of $I(x')$ on $x'$, we can see
immediately that adding a $\delta I_0$
upsets the balance between $I_0$ and $I$,
leading to a net positive or negative accumulation to the integral.

We now consider the most simple case, assuming $I = I_0$ = const
(so that the nominal $b(x)$ = constant as well), and again introduce
$\delta I_0$.   This case can be solved directly for the 
integration error $\delta b$, and we obtain
\begin{equation}
\delta b = K_B \int \frac{\delta I_0}{I_0} \,dx = K_B \frac{\delta I_0}{I_0} \Delta x,
\end{equation}
where $\Delta x$ is the integration distance.
This shows that $\delta I_0$ produces an integration error $\delta b$ which accumulates
in space, with slope $K_B \delta I_0 / I_0$.
This result matches with the numerical example in Fig~\ref{Fig_Multi_I0}(a):  taking $K_B$ = 59~T,
$\delta I_0 / I$ = 10\%, and $\Delta x= 2$~mm, we get an accumulated
$\delta b$ error $\approx 12$~T-mm.  Therefore this explains the 
overall positive and negative slopes introduced by  $\delta I_0$.  Note further features 
will also be introduced in the general case, such as the spatial offsets of where the magnetic jump occurs in Fig.~\ref{Fig_Multi_I0}(a),
produced by the additional non-linearity of the proton mapping.

This result directly connects magnetic field measurement errors to errors in $I_0$.
Similar classes of ``integration errors'' are common in other fields, such as electronic integration circuits,
whenever small but correlated offset errors integrate to produce spurious signals.
This issue illustrates a challenge of reconstructing magnetic fields from
 proton fluence measurements.
 We note this issue generalizes to input
proton fluence \textit{non-uniformities}, i.e. an $x$-dependence of $I_0(x)$.  Such non-uniformities, if not characterized
and accounted for in the analysis, also integrate up to produce spurious
magnetic fields.  This effect is worse at long wavelengths, and for longer integration domains ($\delta b \propto \Delta x$).

\section{Boundary-constrained reconstruction}

\label{Section_BC}

The previous section showed how different $I_0$ can produce significantly different
reconstructions, and specifically how the changes to the average $\bar{I_0}$ leads to different
ramp rates of the magnetic field across the domain.
Accordingly, it is important to constrain this quantity for experimental analysis.
In this section, we develop a \textit{boundary-constrained} reconstruction
which uses a pair of boundary conditions on the magnetic field,
which yields a reconstruction which passes through both specified boundary conditions,
and in the process infers the required average $\bar{I_0}$.

\begin{figure}
\centering
\includegraphics{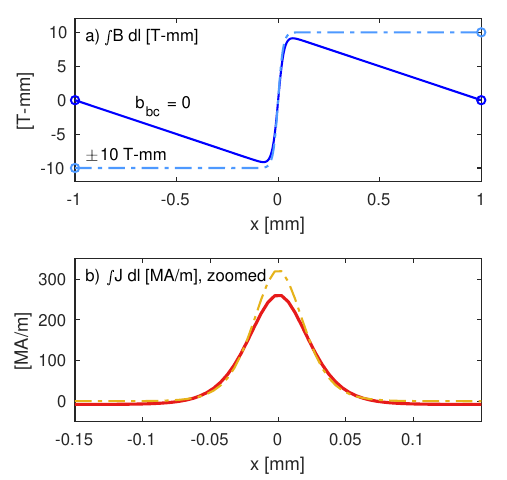}
\caption{Demonstration of radiography reconstructions to achieve specified boundary conditions,
where $\bar{I_0}$ is considered a free parameter.
a) Reconstruction of line-integrated magnetic field to achieve $b_{bc} = 0$ (dark blue),
or $b_{bc}= \pm 10$~T-mm (light blue, dashed) at the two ends of the domain.
b) Current profiles determined for each case, where the red
curve is the $b_{bc} = 0$ case and the yellow
dashed curve is for the  $b_{bc} = \pm 10$~T-mm case.  
}
\label{Fig_BC}
\end{figure}

We recall that Fig.~\ref{Fig_Multi_I0} shows that using different $I_0$ produces
a large family of magnetic field solutions which ramp to nearly
arbitrary values elsewhere in the domain.
However, the converse is also true: \textit{if 
the magnetic field boundary conditions are specified at multiple points, 
then $I_0$ can be chosen so that the
reconstructed field passes through those points.}
There is a relationship between $I_0$
and the boundary conditions on $b(x)$.

We now directly construct this, using the principle of conservation of protons:
imagine that we specify a \textit{two-point} boundary condition on $b(x)$, namely
$b(x_1) = b_1$ and $b(x_2) =  b_2$.   Then
we can calculate  $x'_1 = x_1 +  b_1/K_B$ and $x'_2 = x_2 + b_2/K_B$.
From this we integrate the total
protons landing between $x'_1$ and $x'_2$ which is $\int_{x'_1}^{x'_2} I(x') \,dx'$.
This determines $\bar{I_0}$, which is the average of $I_0$ on $[x_1, x_2]$, via
\begin{equation}
\int_{x'_1}^{x'_2} I(x') \,dx' = \int_{x_1}^{x_2} I_0(x) \,dx \equiv \bar{I_0} \Delta x,
\label{Eq_AvgI0}
\end{equation}
where $\Delta x = x_2-x_1$ is the integration distance.
Therefore, \textit{a two-point boundary condition on the 
magnetic field is sufficient to fix the average 
of the input proton fluence $\bar{I_0}$.}
This relation is exact, even considering the non-linearity of the mapping.

We now implement this new prescription for a final set of reconstructions, shown in Fig.~\ref{Fig_BC},
where we take that $I_0$ is unknown before the analysis, but specify two-point boundary conditions on $b(x)$.
We run two cases, one where we use the
previous $b_{bc} = 0$ at $x_{bc} = \pm 1$~mm, exactly as analyzed in Fig.~\ref{Fig_Reconstruction},
and secondly a case where $(x,b)_{bc}$ are ($-1$~mm, $-10$~T-mm) and
 ($+1$~mm, $+10$~T-mm), which we call the $\pm$10~T-mm case.

The boundary information is used to infer $\bar{I_0}$ for 
each case, via Eq.~\ref{Eq_AvgI0}, which is then used to reconstruct the magnetic fields.
(We again use a uniform $I_0 = \bar{I_0}$ for each case.)
For the $b_{bc}$ = 0 case, $I_0$ is inferred to be exactly 100 protons/pixel (as expected).
For the $b_{bc} = \pm 10$~T-mm case, $I_0$ is inferred to be 120.4 protons/pixel.
The results are shown in Fig.~\ref{Fig_BC}(a),
with the dark-blue curve showing the solution for $b_{bc} = 0$,
and the light-blue dashed curve showing the solution for $b_{bc} = \pm 10$~T-mm.
We observe that the magnetic field exactly
achieves the specified boundary conditions in each case.

%

Finally, Fig.~\ref{Fig_BC}(b) shows a plot
of inferred line-integrated current density $\int J_z\, dz$ for each case,
zoomed to a smaller domain near the current sheet.
We observe the shapes are quantitively similar, but there
is a $\approx 20\%$ difference in the peak current density, due to
the difference in $I_0$ determined for each case.  
The fractional difference in $\int J_z\, dz$ is due to the difference 
$\delta I_0/I_0 \approx 20\%$ between the cases.

To conclude, in this section we developed
a boundary-constrained reconstruction procedure,
which reconstructs a magnetic field profile
matching a pair of magnetic field boundary conditions $(x_1,b_1)$ and $(x_2,b_2)$.
As shown in Fig.~\ref{Fig_BC}, 
given different boundary conditions, 
different reconstructed field profiles can be generated,
even from the same observed proton fluence profile $I(x')$.
The difference comes in through a different inferred (average) source fluence $\bar{I_0}$ for each case.
These results highlight the general importance of
applying boundary condition information for reconstruction analysis, 
and secondly the interrelation between boundary conditions and 
the source proton fluence.

A final point is that the source proton fluence may not be perfectly uniform.
The discussion here shows that the 
magnetic field boundary conditions will constrain the 
average of the input proton fluence.  However, it may still be clear in the raw
data that additional  ``structure'' and spatial dependence of the proton fluence $I_0(x)$
should be accounted for.  In general, additional information will be needed to constrain structure of the source fluence.  
Some statistical strategies for non-uniform fluence profiles have been discussed in Ref.~\cite{KasimPRE2019}.
We leave this to be pursued on a case
by case basis, and for future publications which focus primarily on experimental data.

\section{Reconstruction with experimental data}

\label{Section_NIF}

We conclude with an example reconstruction using data from magnetic reconnection experiments
from the National Ignition Facility \cite{Fox2023}.
The goal of this section is to show a demonstration
reconstruction of raw experimental data and the
importance of boundary conditions.
Here we show a ``hybrid'' 
proton deflectometry scheme with mesh and fluence
reconstruction regions, which allows us
to directly implement the boundary-constrained analysis of Section~\ref{Section_BC} to
complete a reconstruction with experimentally-determined boundary conditions.

In the experiments, two separate plumes are produced by multiple lasers irradiating
a solid carbon target.   The expanding plumes self-magnetize through the Biermann
battery effect, which produces a toroidal field wrapping around each plume \cite{PetrassoPRL2009}.
Subsequent collision of the two plumes compresses the opposite fields from the two plumes 
and drives magnetic reconnection \cite{NilsonPRL2006, LiPRL2007b, FikselPRL2014, RosenbergPRL2015}.
Measuring magnetic fields in the experiments is valuable to
address scientific goals such as understanding 
the structure of the current sheet, growth of instabilities,
and quantities such as the upstream
magnetic fields, plasma current density, 
the width of the current sheet, 
and rate of magnetic reconnection \cite{FoxPRL2011, FoxPoP2012b, LezhninPoP2018} which can be compared against
magnetic reconnection models and observations from space and  astrophysical plasmas \cite{YamadaRMP2010}.

\begin{figure}[h]
\centering
\includegraphics[width=3.375in]{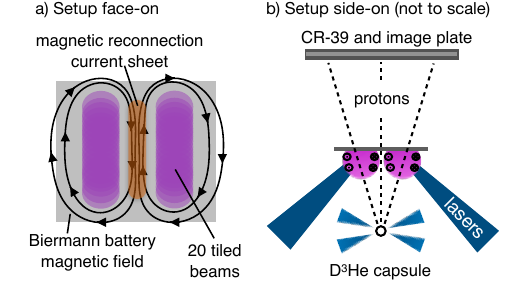}
\includegraphics[width=3.375in]{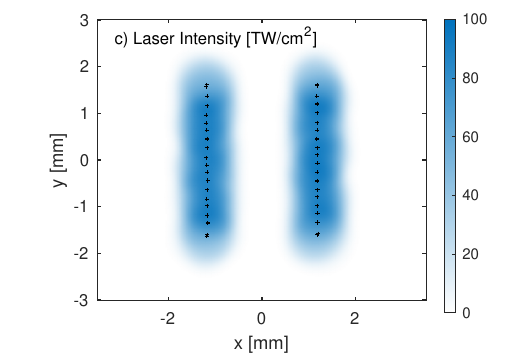}
\caption{Experimental setup at NIF.
a) Schematic of laser profile on-target, Biermann battery
magnetic fields, and current sheet.  b) Schematic
side-on view of target including DHe$^3$ capsule
and detector stack (not to scale).  c) Laser intensity
profile on-target from two sets of 20 tiled laser beams.
Individual laser foci are indicated as the crossed data points.}
\label{Fig_NIF_Setup}
\end{figure}

The experimental geometry is very similar to Fig.~\ref{Fig_Setup}
and is shown schematically in Fig.~\ref{Fig_NIF_Setup}.
To achieve the quasi-1-D geometry, two lines of 20 laser beams were tiled onto a flat target
occupying the $(x,y)$ plane.
The laser foci were in two groups focused at $x=\pm1.2$~mm, 
with each group tiled in the $y$-direction over 4~mm,
producing two highly-elongated plumes.
The two plumes collided at $x=0$, producing
a quasi-1D current sheet.
The peak, overlapped laser intensity on-target was $I_L =$~\sciexp{1}{14}~W/cm$^2$,
with a 0.6~ns square pulse.
The beams were all in the group of ``outer'' beams at NIF which
had on-target angles 44--50 degrees, and
used phase plates designed to focus to a circular
focal spot in the horizontal plane of the target with FWHM 1.24--1.28~mm.
The calculated intensity profile is shown in Fig.~\ref{Fig_NIF_Setup}(c).
A \DHe{} backlighter was imploded by separate drive
beams a distance $L_s = 20$~mm away from the foil.
A detector stack was mounted a distance $L_d=222$~mm on the opposite
side.  The data from 3 and 14.7~MeV protons was registered
on CR-39~proton-track detectors and scanned by standard
techniques~\cite{SeguinRSI2003}.

We attached laser-cut meshes (Au, thickness 76~$\mu$m)
 to the back of the target over some regions to break
the protons into beamlets.
Tracking the beamlet deflections provides a direct magnetic field
measurement at these locations via Eq.~\ref{Eq_mapping}.
In addition to the CR-39 detectors, the detector stack 
also included an image plate (IP), positioned after the 14.7~MeV CR-39 which measured
x-rays produced by the \DHe3{} implosion.
The IP records the x-ray shadow of the mesh,
which provides the absolute reference location for
each beamlet \cite{JohnsonRSI2022, MalkoApplOptics2022}.
Filtering through the stack provides a low-energy cutoff of $h\nu \gtrsim 25$~keV for
the x-rays reaching the IP.
The image plate and CR-39's were co-aligned to better than 0.5 pixel in post-processing analysis using the
toothed features at the image boundary, imprinted by a fiducial frame at the
front of the detector stack.

\begin{figure}[h!]
\centering
\includegraphics[width=3.375in]{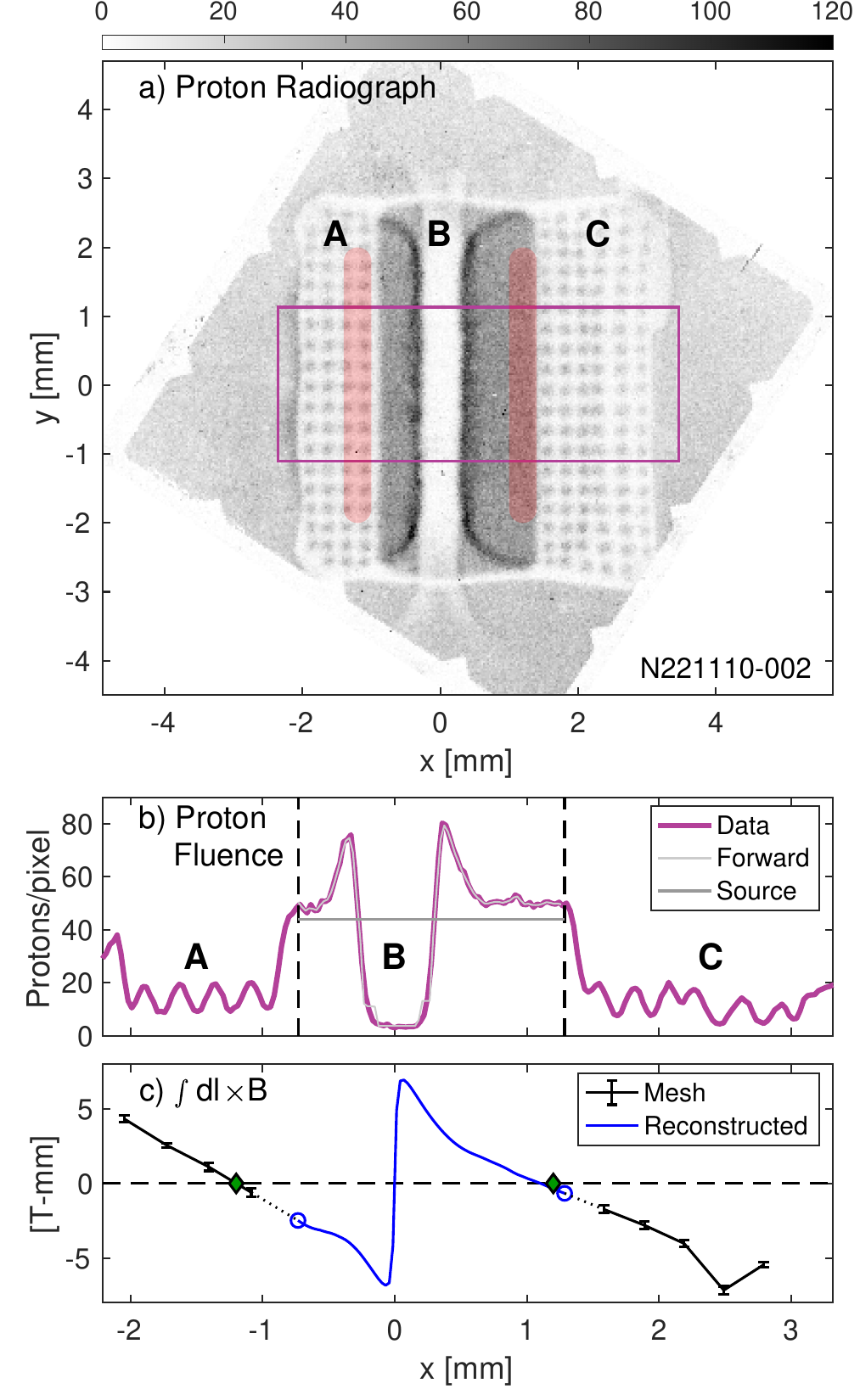}
\caption{Proton radiography data from magnetic reconnection
experiments at NIF and associated magnetic field reconstructions.
a) Raw proton radiograph image, containing both mesh regions `A' and `C' and
a mesh-free region `B' for fluence analysis.  The footprints of the 
drive laser groups are indicated as the light red ovals.
b) Profile of the proton fluence data (purple) over the purple box from (a), averaged
in the $y$-direction.  These data were used to reconstruct the 
field in panel (c).  The dark gray line indicates the inferred source proton fluence, 
and the light gray line indicates the forward model
proton fluence from the reconstruction.
c)  Direct measurements of $\int B_y\,dz$ (black points with error bars) from 
mesh data, and reconstructed $\int B_y\,dz$ from the proton 
fluence data (blue).  Boundary conditions for the reconstruction are indicated by the 
blue circles at the ends of the reconstructed curve, and were
determined from extrapolating the mesh data.   Green diamonds indicate
the centers of the laser-foci at $x = \pm 1.2$~mm where $\int B_y\,dz =0$ is expected
by symmetry.
} 
\label{Fig_NIF}
\end{figure}

Figure~\ref{Fig_NIF}(a) shows a raw 14.7~MeV proton radiograph from
the experiment.  The backlighter was timed such that the 14.7~MeV protons
crossed the plasma plane 3.0~ns after the start of the main laser drive. 
Meshes (indicated by `A' and `C'),
were used on the non-reconnecting sides of the plumes, while
the mesh-free region (`B') in the reconnection region 
allows for a magnetic field reconstruction.
The footprints of the two groups of drive lasers are indicated by the red ovals
centered at $x = \pm 1.2$~mm, which are extended along $y$ by tiling the lasers,
and denote the centroid of each plume.
Two high-proton-fluence (dark) bands,
visible in region B, are produced  
 by the toroidal Biermann-battery fields wrapping around each plume.
The two plumes collide to produce a reconnection current 
sheet which is observed as a low-fluence (light) region centered at $x=0$.
 The magnetic field in the current sheet region is quasi-1-D, 
allowing the 1-D analysis developed in this paper.

The proton fluence profile along $x$ is plotted in Fig.~\ref{Fig_NIF}(b),
and shown as the purple curve.
In regions A and C, the mesh modulation is visible.
Region B provides the proton fluence data from the central current sheet region.
The fluence profile plotted in Fig.~\ref{Fig_NIF}(b) is obtained by averaging over 
$y \in [-1,1]$~mm.
The raw fluence data shows qualitative agreement with the 
analytic profiles in Fig.~\ref{Fig_Analytic}(d), especially the 
broad fluence depletion near $x = 0$, indicating the 
current sheet where the magnetic field rapidly reverses.

Figure~\ref{Fig_NIF}(c) shows the magnetic field reconstructed from this
fluence data.     We apply the boundary-constrained
reconstruction developed in Section~\ref{Section_BC}.  
The boundary conditions for the magnetic field were
obtained from the mesh data.  
The deflection of each beamlet from its reference location
was obtained by comparing the beamlet data between
the 14.7~MeV proton and the image plate x-ray data \cite{JohnsonRSI2022, MalkoApplOptics2022}.
This directly provides the magnetic field at each beamlet, plotted 
as the black data points in Fig.~\ref{Fig_NIF}(c),
where the error bars indicate the standard deviation within each column.
Since the first mesh
cell is some distance into the mesh region, we extrapolated
the boundary condition linearly from the two nearest
mesh points to the first available point in the fluence region.   These points are shown
in Fig.~\ref{Fig_NIF}(c) as the closed blue circles
at the ends of the magnetic field reconstruction.
The magnetic field is then reconstructed
from the fluence data, shown as the 
blue curve.
The raw proton fluence did not show
strong evidence that the source' proton
fluence was non-uniform, so we reconstructed with 
a uniform source fluence, with the fluence level inferred from the
boundary-constraint analysis from Section~\ref{Section_BC},
shown as the light gray line in Fig.~\ref{Fig_NIF}(b).
The magnetic field profile has a qualitatively
similar shape to the prior analytic examples,
with the peak line-integrated magnetic field strength near 7~T-mm. 
The nominal centers of the laser foci
for each plume are indicated as green diamonds
 at $x= \pm{1.2}$~mm, and we observe that the inferred
magnetic field crosses through zero near these points, as expected by the 
symmetry of the Biermann-battery field-generation process.

As a final check, the reconstructed magnetic field profile was used to generate 
the forward-model proton fluence, plotted as
the thin gray curve in Fig.~\ref{Fig_NIF}(b), 
which is seen to be in good agreement
with the raw proton data.   Some fine-scale jaggedness in the data
is observed in the current sheet, which is a result of the finite spatial resolution 
of the proton data (equivalent to 28~$\mu$m in the plasma plane).
We always apply this 
check to verify a minimum of 
agreement between the inferred magnetic 
field and the data.

\section{Discussion and Conclusions}

A 1-D reconstruction procedure has been developed to
infer electromagnetic fields in high-energy-density plasma
experiments.  A 1-D reconstruction is valuable 
as a complement and cross-check to 2-D analyses and may be directly applied to
1-D experiments.  Here we verified the 1-D reconstruction algorithm against semi-analytical field models,
and presented an example showing it can reconstruct
magnetic field data from a recent experiment at NIF.
A software package PRADICAMENT was developed
which implements this algorithm and is discussed in the Appendix. 

Using the algorithm we then explored
the relation between the boundary conditions and source fluence
for 1-D reconstructions.
First, we showed that proton fluence measurements 
are closely related to plasma current measurements, 
which is why an integration is required to infer the magnetic fields from the proton fluence.
A boundary condition is needed to obtain the absolute magnetic field,
otherwise the analysis produces only the relative
change of the magnetic field across the domain.
Next, we showed that if there are errors or uncertainties in characterizing source
proton fluence, then this leads to ``integration errors''
in reconstructing the magnetic field, which increases with the length of the reconstructed domain.
However, we showed that applying \textit{two-point} boundary conditions on the magnetic field
 ($x_1, b_1$), and ($x_2, b_2$),
 can be used to fix the average source proton
fluence on the interval $[x_1, x_2]$, and from this we developed a boundary-constrained
reconstruction in Section~\ref{Section_BC}.

From these considerations, we conclude that 
boundary conditions can be critical input data to the analysis.
It is obviously a best practice to directly confirm the
boundary conditions with experimental measurements.
To do so we developed a ``hybrid'' proton 
deflectometry technique, which combines
fluence analysis with boundary conditions from beamlet mapping 
\cite{JohnsonRSI2022, MalkoApplOptics2022}.
The direct magnetic field measurements from the
beamlet analysis were used as boundary conditions to constrain a 
high-resolution fluence reconstruction in Section~\ref{Section_NIF}.

The present 1-D algorithm has some valuable complementary features compared to 
Monge-Ampere-type reconstruction implementations (e.g. \cite{BottJPP2017}).
While not developed extensively here, 
it is also possible to run the present
algorithm in a ``free-boundary'' mode.  
This mode takes as input the observed fluence $I(x')$ 
and a specified source fluence $I_0(x)$, 
and produces a reconstructed magnetic field.  
With no boundary conditions applied, one should remember this mode only produces the 
relative change of the magnetic field over the domain, and 
one should test the sensitivity of the results to uncertainty in $I_0$.
Nevertheless this technique could also be useful under certain circumstances,
and especially for ``short-baseline'' reconstructions where there is not a long
distance for integration errors to pile up; in contrast, the
Monge-Ampere type relaxation solvers always produce a magnetic field with
fixed boundary conditions (with commonly-used reconstruction codes
using $\mathbf{B_{\rm tangential}} = 0$, 
as of the time of this writing \cite{PROBLEM_2024}).

Since the goal of this work has been to demonstrate the 
overall reconstruction procedure, we defer detailed error analysis to future 
publications, but one can imagine that several sources of error can be
incorporated, including the uncertainty in boundary conditions, 
finite spatial resolution effects, fluence background, and
non-uniformity in the source proton fluence.   For the present
data it appeared that the source proton fluence was relatively flat,
so we used a flat source fluence profile for the simplest possible reconstruction;
however future work will consider constraining non-uniformity
in the source fluence as part of the analysis, and how this feeds 
into uncertainty in the final reconstructed magnetic field.  
Once again, having the ground truth of direct magnetic field data at several points
will form an important constraint on the final profiles.

\acknowledgments

This work was supported by DOE Award No. DE-NA-004034.
We acknowledge experimental data provided by the NIF Discovery Science Program, supported by the U.S. DOE Office of Fusion Energy Sciences under FWP SW1626 FES.   J.G.-M. was supported by 
the National Science Foundation Graduate Research Fellowship Program under Grant No. 2039656.

\bibliography{refs}

\begin{thebibliography}{32}%
\makeatletter
\providecommand \@ifxundefined [1]{%
 \@ifx{#1\undefined}
}%
\providecommand \@ifnum [1]{%
 \ifnum #1\expandafter \@firstoftwo
 \else \expandafter \@secondoftwo
 \fi
}%
\providecommand \@ifx [1]{%
 \ifx #1\expandafter \@firstoftwo
 \else \expandafter \@secondoftwo
 \fi
}%
\providecommand \natexlab [1]{#1}%
\providecommand \enquote  [1]{``#1''}%
\providecommand \bibnamefont  [1]{#1}%
\providecommand \bibfnamefont [1]{#1}%
\providecommand \citenamefont [1]{#1}%
\providecommand \href@noop [0]{\@secondoftwo}%
\providecommand \href [0]{\begingroup \@sanitize@url \@href}%
\providecommand \@href[1]{\@@startlink{#1}\@@href}%
\providecommand \@@href[1]{\endgroup#1\@@endlink}%
\providecommand \@sanitize@url [0]{\catcode `\\12\catcode `\$12\catcode
  `\&12\catcode `\#12\catcode `\^12\catcode `\_12\catcode `\%12\relax}%
\providecommand \@@startlink[1]{}%
\providecommand \@@endlink[0]{}%
\providecommand \url  [0]{\begingroup\@sanitize@url \@url }%
\providecommand \@url [1]{\endgroup\@href {#1}{\urlprefix }}%
\providecommand \urlprefix  [0]{URL }%
\providecommand \Eprint [0]{\href }%
\providecommand \doibase [0]{https://doi.org/}%
\providecommand \selectlanguage [0]{\@gobble}%
\providecommand \bibinfo  [0]{\@secondoftwo}%
\providecommand \bibfield  [0]{\@secondoftwo}%
\providecommand \translation [1]{[#1]}%
\providecommand \BibitemOpen [0]{}%
\providecommand \bibitemStop [0]{}%
\providecommand \bibitemNoStop [0]{.\EOS\space}%
\providecommand \EOS [0]{\spacefactor3000\relax}%
\providecommand \BibitemShut  [1]{\csname bibitem#1\endcsname}%
\let\auto@bib@innerbib\@empty
\bibitem [{\citenamefont {Kugland}\ \emph {et~al.}(2012)\citenamefont
  {Kugland}, \citenamefont {Ryutov}, \citenamefont {Plechaty}, \citenamefont
  {Ross},\ and\ \citenamefont {Park}}]{KuglandRSI2012}%
  \BibitemOpen
  \bibfield  {author} {\bibinfo {author} {\bibfnamefont {N.~L.}\ \bibnamefont
  {Kugland}}, \bibinfo {author} {\bibfnamefont {D.~D.}\ \bibnamefont {Ryutov}},
  \bibinfo {author} {\bibfnamefont {C.}~\bibnamefont {Plechaty}}, \bibinfo
  {author} {\bibfnamefont {J.~S.}\ \bibnamefont {Ross}},\ and\ \bibinfo
  {author} {\bibfnamefont {H.~S.}\ \bibnamefont {Park}},\ }\bibfield  {title}
  {\bibinfo {title} {{Invited Article: Relation between electric and magnetic
  field structures and their proton-beam images}},\ }\href
  {https://doi.org/10.1063/1.4750234} {\bibfield  {journal} {\bibinfo
  {journal} {Rev.\ Sci.\ Inst.}\ }\textbf {\bibinfo {volume}
  {83}},\ \bibinfo {pages} {101301} (\bibinfo {year} {2012})}\BibitemShut
  {NoStop}%
\bibitem [{\citenamefont {Bott}\ \emph {et~al.}(2017)\citenamefont {Bott},
  \citenamefont {Graziani}, \citenamefont {Tzeferacos}, \citenamefont {White},
  \citenamefont {Lamb}, \citenamefont {Gregori},\ and\ \citenamefont
  {Schekochihin}}]{BottJPP2017}%
  \BibitemOpen
  \bibfield  {author} {\bibinfo {author} {\bibfnamefont {A.~F.~A.}\
  \bibnamefont {Bott}}, \bibinfo {author} {\bibfnamefont {C.}~\bibnamefont
  {Graziani}}, \bibinfo {author} {\bibfnamefont {P.}~\bibnamefont
  {Tzeferacos}}, \bibinfo {author} {\bibfnamefont {T.~G.}\ \bibnamefont
  {White}}, \bibinfo {author} {\bibfnamefont {D.~Q.}\ \bibnamefont {Lamb}},
  \bibinfo {author} {\bibfnamefont {G.}~\bibnamefont {Gregori}},\ and\ \bibinfo
  {author} {\bibfnamefont {A.~A.}\ \bibnamefont {Schekochihin}},\ }\bibfield
  {title} {\bibinfo {title} {Proton imaging of stochastic magnetic fields},\
  }\bibfield  {journal} {\bibinfo  {journal} {J. Plasma Phys.}\
  }
  (\bibinfo {year} {2017})\BibitemShut {NoStop}%
\bibitem [{\citenamefont {Schaeffer}\ \emph {et~al.}(2023)\citenamefont
  {Schaeffer}, \citenamefont {Bott}, \citenamefont {Borghesi}, \citenamefont
  {Flippo}, \citenamefont {Fox}, \citenamefont {Fuchs}, \citenamefont {Li},
  \citenamefont {S{\'e}guin}, \citenamefont {Park}, \citenamefont
  {Tzeferacos},\ and\ \citenamefont {Willingale}}]{SchaefferRMP2023}%
  \BibitemOpen
  \bibfield  {author} {\bibinfo {author} {\bibfnamefont {D.~B.}\ \bibnamefont
  {Schaeffer}}, \bibinfo {author} {\bibfnamefont {A.~F.~A.}\ \bibnamefont
  {Bott}}, \bibinfo {author} {\bibfnamefont {M.}~\bibnamefont {Borghesi}},
  \bibinfo {author} {\bibfnamefont {K.~A.}\ \bibnamefont {Flippo}}, \bibinfo
  {author} {\bibfnamefont {W.}~\bibnamefont {Fox}}, \bibinfo {author}
  {\bibfnamefont {J.}~\bibnamefont {Fuchs}}, \bibinfo {author} {\bibfnamefont
  {C.}~\bibnamefont {Li}}, \bibinfo {author} {\bibfnamefont {F.~H.}\
  \bibnamefont {S{\'e}guin}}, \bibinfo {author} {\bibfnamefont {H.-S.}\
  \bibnamefont {Park}}, \bibinfo {author} {\bibfnamefont {P.}~\bibnamefont
  {Tzeferacos}},\ and\ \bibinfo {author} {\bibfnamefont {L.}~\bibnamefont
  {Willingale}},\ }\bibfield  {title} {\bibinfo {title} {Proton imaging of
  high-energy-density laboratory plasmas},\ }\href
  {https://doi.org/10.1103/revmodphys.95.045007} {\bibfield  {journal}
  {\bibinfo  {journal} {Rev.\ Mod.\ Phys.}\ }\textbf {\bibinfo {volume}
  {95}},\ \bibinfo {pages} {045007} (\bibinfo {year} {2023})}\BibitemShut
  {NoStop}%
\bibitem [{\citenamefont {Gotchev}\ \emph {et~al.}(2009)\citenamefont
  {Gotchev}, \citenamefont {Chang}, \citenamefont {Knauer}, \citenamefont
  {Meyerhofer}, \citenamefont {Polomarov}, \citenamefont {Frenje},
  \citenamefont {Li}, \citenamefont {Manuel}, \citenamefont {Petrasso},
  \citenamefont {Rygg}, \citenamefont {S{\'e}guin},\ and\ \citenamefont
  {Betti}}]{GotchevPRL2009}%
  \BibitemOpen
  \bibfield  {author} {\bibinfo {author} {\bibfnamefont {O.~V.}\ \bibnamefont
  {Gotchev}}, \bibinfo {author} {\bibfnamefont {P.~Y.}\ \bibnamefont {Chang}},
  \bibinfo {author} {\bibfnamefont {J.~P.}\ \bibnamefont {Knauer}}, \bibinfo
  {author} {\bibfnamefont {D.~D.}\ \bibnamefont {Meyerhofer}}, \bibinfo
  {author} {\bibfnamefont {O.}~\bibnamefont {Polomarov}}, \bibinfo {author}
  {\bibfnamefont {J.}~\bibnamefont {Frenje}}, \bibinfo {author} {\bibfnamefont
  {C.~K.}\ \bibnamefont {Li}}, \bibinfo {author} {\bibfnamefont {M.~J.-E.}\
  \bibnamefont {Manuel}}, \bibinfo {author} {\bibfnamefont {R.~D.}\
  \bibnamefont {Petrasso}}, \bibinfo {author} {\bibfnamefont {J.~R.}\
  \bibnamefont {Rygg}}, \bibinfo {author} {\bibfnamefont {F.~H.}\ \bibnamefont
  {S{\'e}guin}},\ and\ \bibinfo {author} {\bibfnamefont {R.}~\bibnamefont
  {Betti}},\ }\bibfield  {title} {\bibinfo {title} {Laser-driven magnetic-flux
  compression in high-energy-density plasmas},\ }\href
  {https://doi.org/10.1103/physrevlett.103.215004} {\bibfield  {journal}
  {\bibinfo  {journal} {Phys.\ Rev.\ Lett.}\ }\textbf {\bibinfo {volume}
  {103}},\ \bibinfo {pages} {215004} (\bibinfo {year} {2009})}\BibitemShut
  {NoStop}%
\bibitem [{\citenamefont {Li}\ \emph {et~al.}(2006)\citenamefont {Li},
  \citenamefont {S\'{e}guin}, \citenamefont {Frenje}, \citenamefont {Rygg},
  \citenamefont {Petrasso}, \citenamefont {Town}, \citenamefont {Amendt},
  \citenamefont {Hatchett}, \citenamefont {Landen}, \citenamefont {Mackinnon},
  \citenamefont {Patel}, \citenamefont {Smalyuk}, \citenamefont {Sangster},\
  and\ \citenamefont {Knauer}}]{LiPRL2006}%
  \BibitemOpen
  \bibfield  {author} {\bibinfo {author} {\bibfnamefont {C.~K.}\ \bibnamefont
  {Li}}, \bibinfo {author} {\bibfnamefont {F.~H.}\ \bibnamefont {S\'{e}guin}},
  \bibinfo {author} {\bibfnamefont {J.~A.}\ \bibnamefont {Frenje}}, \bibinfo
  {author} {\bibfnamefont {J.~R.}\ \bibnamefont {Rygg}}, \bibinfo {author}
  {\bibfnamefont {R.~D.}\ \bibnamefont {Petrasso}}, \bibinfo {author}
  {\bibfnamefont {R.~P.~J.}\ \bibnamefont {Town}}, \bibinfo {author}
  {\bibfnamefont {P.~A.}\ \bibnamefont {Amendt}}, \bibinfo {author}
  {\bibfnamefont {S.~P.}\ \bibnamefont {Hatchett}}, \bibinfo {author}
  {\bibfnamefont {O.~L.}\ \bibnamefont {Landen}}, \bibinfo {author}
  {\bibfnamefont {A.~J.}\ \bibnamefont {Mackinnon}}, \bibinfo {author}
  {\bibfnamefont {P.~K.}\ \bibnamefont {Patel}}, \bibinfo {author}
  {\bibfnamefont {V.~A.}\ \bibnamefont {Smalyuk}}, \bibinfo {author}
  {\bibfnamefont {T.~C.}\ \bibnamefont {Sangster}},\ and\ \bibinfo {author}
  {\bibfnamefont {J.~P.}\ \bibnamefont {Knauer}},\ }\bibfield  {title}
  {\bibinfo {title} {{Measuring \$E\$ and \$B\$ Fields in Laser-Produced
  Plasmas with Monoenergetic Proton Radiography}},\ }\href
  {https://doi.org/10.1103/physrevlett.97.135003} {\bibfield  {journal}
  {\bibinfo  {journal} {Phys.\ Rev.\ Lett.}\ }\textbf {\bibinfo {volume}
  {97}},\ \bibinfo {pages} {135003} (\bibinfo {year} {2006})}\BibitemShut
  {NoStop}%
\bibitem [{\citenamefont {Petrasso}\ \emph {et~al.}(2009)\citenamefont
  {Petrasso}, \citenamefont {Li}, \citenamefont {Seguin}, \citenamefont {Rygg},
  \citenamefont {Frenje}, \citenamefont {Betti}, \citenamefont {Knauer},
  \citenamefont {Meyerhofer}, \citenamefont {Amendt}, \citenamefont {Froula},
  \citenamefont {Landen}, \citenamefont {Patel}, \citenamefont {Ross},\ and\
  \citenamefont {Town}}]{PetrassoPRL2009}%
  \BibitemOpen
  \bibfield  {author} {\bibinfo {author} {\bibfnamefont {R.~D.}\ \bibnamefont
  {Petrasso}}, \bibinfo {author} {\bibfnamefont {C.~K.}\ \bibnamefont {Li}},
  \bibinfo {author} {\bibfnamefont {F.~H.}\ \bibnamefont {Seguin}}, \bibinfo
  {author} {\bibfnamefont {J.~R.}\ \bibnamefont {Rygg}}, \bibinfo {author}
  {\bibfnamefont {J.~A.}\ \bibnamefont {Frenje}}, \bibinfo {author}
  {\bibfnamefont {R.}~\bibnamefont {Betti}}, \bibinfo {author} {\bibfnamefont
  {J.~P.}\ \bibnamefont {Knauer}}, \bibinfo {author} {\bibfnamefont {D.~D.}\
  \bibnamefont {Meyerhofer}}, \bibinfo {author} {\bibfnamefont {P.~A.}\
  \bibnamefont {Amendt}}, \bibinfo {author} {\bibfnamefont {D.~H.}\
  \bibnamefont {Froula}}, \bibinfo {author} {\bibfnamefont {O.~L.}\
  \bibnamefont {Landen}}, \bibinfo {author} {\bibfnamefont {P.~K.}\
  \bibnamefont {Patel}}, \bibinfo {author} {\bibfnamefont {J.~S.}\ \bibnamefont
  {Ross}},\ and\ \bibinfo {author} {\bibfnamefont {R.~P.~J.}\ \bibnamefont
  {Town}},\ }\bibfield  {title} {\bibinfo {title} {{Lorentz Mapping of Magnetic
  Fields in Hot Dense Plasmas}},\ }\href
  {https://doi.org/10.1103/physrevlett.103.085001} {\bibfield  {journal}
  {\bibinfo  {journal} {Phys. Rev. Lett.}\ }\textbf {\bibinfo {volume} {103}},\
  \bibinfo {pages} {085001} (\bibinfo {year} {2009})}\BibitemShut {NoStop}%
\bibitem [{\citenamefont {Gao}\ \emph {et~al.}(2015)\citenamefont {Gao},
  \citenamefont {Nilson}, \citenamefont {Igumenshchev}, \citenamefont {Haines},
  \citenamefont {Froula}, \citenamefont {Betti},\ and\ \citenamefont
  {Meyerhofer}}]{GaoPRL2015}%
  \BibitemOpen
  \bibfield  {author} {\bibinfo {author} {\bibfnamefont {L.}~\bibnamefont
  {Gao}}, \bibinfo {author} {\bibfnamefont {P.~M.}\ \bibnamefont {Nilson}},
  \bibinfo {author} {\bibfnamefont {I.~V.}\ \bibnamefont {Igumenshchev}},
  \bibinfo {author} {\bibfnamefont {M.~G.}\ \bibnamefont {Haines}}, \bibinfo
  {author} {\bibfnamefont {D.~H.}\ \bibnamefont {Froula}}, \bibinfo {author}
  {\bibfnamefont {R.}~\bibnamefont {Betti}},\ and\ \bibinfo {author}
  {\bibfnamefont {D.~D.}\ \bibnamefont {Meyerhofer}},\ }\bibfield  {title}
  {\bibinfo {title} {{Precision Mapping of Laser-Driven Magnetic Fields and
  Their Evolution in High-Energy-Density Plasmas}},\ }\href
  {https://doi.org/10.1103/physrevlett.114.215003} {\bibfield  {journal}
  {\bibinfo  {journal} {Phys.\ Rev.\ Lett.}\ }\textbf {\bibinfo {volume}
  {114}},\ \bibinfo {pages} {215003} (\bibinfo {year} {2015})}\BibitemShut
  {NoStop}%
\bibitem [{\citenamefont {Fox}\ \emph {et~al.}(2013)\citenamefont {Fox},
  \citenamefont {Fiksel}, \citenamefont {Bhattacharjee}, \citenamefont {Chang},
  \citenamefont {Germaschewski}, \citenamefont {Hu},\ and\ \citenamefont
  {Nilson}}]{FoxPRL2013}%
  \BibitemOpen
  \bibfield  {author} {\bibinfo {author} {\bibfnamefont {W.}~\bibnamefont
  {Fox}}, \bibinfo {author} {\bibfnamefont {G.}~\bibnamefont {Fiksel}},
  \bibinfo {author} {\bibfnamefont {A.}~\bibnamefont {Bhattacharjee}}, \bibinfo
  {author} {\bibfnamefont {P.~Y.}\ \bibnamefont {Chang}}, \bibinfo {author}
  {\bibfnamefont {K.}~\bibnamefont {Germaschewski}}, \bibinfo {author}
  {\bibfnamefont {S.~X.}\ \bibnamefont {Hu}},\ and\ \bibinfo {author}
  {\bibfnamefont {P.~M.}\ \bibnamefont {Nilson}},\ }\bibfield  {title}
  {\bibinfo {title} {{Filamentation Instability of Counterstreaming
  Laser-Driven Plasmas}},\ }\href
  {https://doi.org/10.1103/physrevlett.111.225002} {\bibfield  {journal}
  {\bibinfo  {journal} {Phys.\ Rev.\ Lett.}\ }\textbf {\bibinfo {volume}
  {111}},\ \bibinfo {pages} {225002} (\bibinfo {year} {2013})}\BibitemShut
  {NoStop}%
\bibitem [{\citenamefont {Nilson}\ \emph {et~al.}(2006)\citenamefont {Nilson},
  \citenamefont {Willingale}, \citenamefont {Kaluza}, \citenamefont
  {Kamperidis}, \citenamefont {Minardi}, \citenamefont {Wei}, \citenamefont
  {Fernandes}, \citenamefont {Notley}, \citenamefont {Bandyopadhyay},
  \citenamefont {Sherlock}, \citenamefont {Kingham}, \citenamefont {Tatarakis},
  \citenamefont {Najmudin}, \citenamefont {Rozmus}, \citenamefont {Evans},
  \citenamefont {Haines}, \citenamefont {Dangor},\ and\ \citenamefont
  {Krushelnick}}]{NilsonPRL2006}%
  \BibitemOpen
  \bibfield  {author} {\bibinfo {author} {\bibfnamefont {P.~M.}\ \bibnamefont
  {Nilson}}, \bibinfo {author} {\bibfnamefont {L.}~\bibnamefont {Willingale}},
  \bibinfo {author} {\bibfnamefont {M.~C.}\ \bibnamefont {Kaluza}}, \bibinfo
  {author} {\bibfnamefont {C.}~\bibnamefont {Kamperidis}}, \bibinfo {author}
  {\bibfnamefont {S.}~\bibnamefont {Minardi}}, \bibinfo {author} {\bibfnamefont
  {M.~S.}\ \bibnamefont {Wei}}, \bibinfo {author} {\bibfnamefont
  {P.}~\bibnamefont {Fernandes}}, \bibinfo {author} {\bibfnamefont
  {M.}~\bibnamefont {Notley}}, \bibinfo {author} {\bibfnamefont
  {S.}~\bibnamefont {Bandyopadhyay}}, \bibinfo {author} {\bibfnamefont
  {M.}~\bibnamefont {Sherlock}}, \bibinfo {author} {\bibfnamefont {R.~J.}\
  \bibnamefont {Kingham}}, \bibinfo {author} {\bibfnamefont {M.}~\bibnamefont
  {Tatarakis}}, \bibinfo {author} {\bibfnamefont {Z.}~\bibnamefont {Najmudin}},
  \bibinfo {author} {\bibfnamefont {W.}~\bibnamefont {Rozmus}}, \bibinfo
  {author} {\bibfnamefont {R.~G.}\ \bibnamefont {Evans}}, \bibinfo {author}
  {\bibfnamefont {M.~G.}\ \bibnamefont {Haines}}, \bibinfo {author}
  {\bibfnamefont {A.~E.}\ \bibnamefont {Dangor}},\ and\ \bibinfo {author}
  {\bibfnamefont {K.}~\bibnamefont {Krushelnick}},\ }\bibfield  {title}
  {\bibinfo {title} {{Magnetic Reconnection and Plasma Dynamics in Two-Beam
  Laser-Solid Interactions}},\ }\href
  {https://doi.org/10.1103/physrevlett.97.255001} {\bibfield  {journal}
  {\bibinfo  {journal} {Phys.\ Rev.\ Lett.}\ }\textbf {\bibinfo {volume}
  {97}},\ \bibinfo {pages} {255001} (\bibinfo {year} {2006})}\BibitemShut
  {NoStop}%
\bibitem [{\citenamefont {Li}\ \emph {et~al.}(2007)\citenamefont {Li},
  \citenamefont {S\'{e}guin}, \citenamefont {Frenje}, \citenamefont {Rygg},
  \citenamefont {Petrasso}, \citenamefont {Town}, \citenamefont {Landen},
  \citenamefont {Knauer},\ and\ \citenamefont {Smalyuk}}]{LiPRL2007b}%
  \BibitemOpen
  \bibfield  {author} {\bibinfo {author} {\bibfnamefont {C.~K.}\ \bibnamefont
  {Li}}, \bibinfo {author} {\bibfnamefont {F.~H.}\ \bibnamefont {S\'{e}guin}},
  \bibinfo {author} {\bibfnamefont {J.~A.}\ \bibnamefont {Frenje}}, \bibinfo
  {author} {\bibfnamefont {J.~R.}\ \bibnamefont {Rygg}}, \bibinfo {author}
  {\bibfnamefont {R.~D.}\ \bibnamefont {Petrasso}}, \bibinfo {author}
  {\bibfnamefont {R.~P.~J.}\ \bibnamefont {Town}}, \bibinfo {author}
  {\bibfnamefont {O.~L.}\ \bibnamefont {Landen}}, \bibinfo {author}
  {\bibfnamefont {J.~P.}\ \bibnamefont {Knauer}},\ and\ \bibinfo {author}
  {\bibfnamefont {V.~A.}\ \bibnamefont {Smalyuk}},\ }\bibfield  {title}
  {\bibinfo {title} {{Observation of Megagauss-Field Topology Changes due to
  Magnetic Reconnection in Laser-Produced Plasmas}},\ }\href
  {https://doi.org/10.1103/physrevlett.99.055001} {\bibfield  {journal}
  {\bibinfo  {journal} {Phys.\ Rev.\ Lett.}\ }\textbf {\bibinfo {volume}
  {99}},\ \bibinfo {pages} {055001} (\bibinfo {year} {2007})}\BibitemShut
  {NoStop}%
\bibitem [{\citenamefont {Fiksel}\ \emph {et~al.}(2014)\citenamefont {Fiksel},
  \citenamefont {Fox}, \citenamefont {Bhattacharjee}, \citenamefont {Barnak},
  \citenamefont {Chang}, \citenamefont {Germaschewski}, \citenamefont {Hu},\
  and\ \citenamefont {Nilson}}]{FikselPRL2014}%
  \BibitemOpen
  \bibfield  {author} {\bibinfo {author} {\bibfnamefont {G.}~\bibnamefont
  {Fiksel}}, \bibinfo {author} {\bibfnamefont {W.}~\bibnamefont {Fox}},
  \bibinfo {author} {\bibfnamefont {A.}~\bibnamefont {Bhattacharjee}}, \bibinfo
  {author} {\bibfnamefont {D.~H.}\ \bibnamefont {Barnak}}, \bibinfo {author}
  {\bibfnamefont {P.~Y.}\ \bibnamefont {Chang}}, \bibinfo {author}
  {\bibfnamefont {K.}~\bibnamefont {Germaschewski}}, \bibinfo {author}
  {\bibfnamefont {S.~X.}\ \bibnamefont {Hu}},\ and\ \bibinfo {author}
  {\bibfnamefont {P.~M.}\ \bibnamefont {Nilson}},\ }\bibfield  {title}
  {\bibinfo {title} {{Magnetic Reconnection between Colliding Magnetized
  Laser-Produced Plasma Plumes}},\ }\href
  {https://doi.org/10.1103/physrevlett.113.105003} {\bibfield  {journal}
  {\bibinfo  {journal} {Phys.\ Rev.\ Lett.}\ }\textbf {\bibinfo {volume}
  {113}},\ \bibinfo {pages} {105003} (\bibinfo {year} {2014})}\BibitemShut
  {NoStop}%
\bibitem [{\citenamefont {Rosenberg}\ \emph {et~al.}(2015)\citenamefont
  {Rosenberg}, \citenamefont {Li}, \citenamefont {Fox}, \citenamefont
  {Zylstra}, \citenamefont {Stoeckl}, \citenamefont {S\'{e}guin}, \citenamefont
  {Frenje},\ and\ \citenamefont {Petrasso}}]{RosenbergPRL2015}%
  \BibitemOpen
  \bibfield  {author} {\bibinfo {author} {\bibfnamefont {M.~J.}\ \bibnamefont
  {Rosenberg}}, \bibinfo {author} {\bibfnamefont {C.~K.}\ \bibnamefont {Li}},
  \bibinfo {author} {\bibfnamefont {W.}~\bibnamefont {Fox}}, \bibinfo {author}
  {\bibfnamefont {A.~B.}\ \bibnamefont {Zylstra}}, \bibinfo {author}
  {\bibfnamefont {C.}~\bibnamefont {Stoeckl}}, \bibinfo {author} {\bibfnamefont
  {F.~H.}\ \bibnamefont {S\'{e}guin}}, \bibinfo {author} {\bibfnamefont
  {J.~A.}\ \bibnamefont {Frenje}},\ and\ \bibinfo {author} {\bibfnamefont
  {R.~D.}\ \bibnamefont {Petrasso}},\ }\bibfield  {title} {\bibinfo {title}
  {{Slowing of Magnetic Reconnection Concurrent with Weakening Plasma Inflows
  and Increasing Collisionality in Strongly Driven Laser-Plasma Experiments}},\
  }\href {https://doi.org/10.1103/physrevlett.114.205004} {\bibfield  {journal}
  {\bibinfo  {journal} {Phys. Rev. Lett.}\ }\textbf {\bibinfo {volume} {114}},\
  \bibinfo {pages} {205004} (\bibinfo {year} {2015})}\BibitemShut {NoStop}%
\bibitem [{\citenamefont {Schaeffer}\ \emph {et~al.}(2019)\citenamefont
  {Schaeffer}, \citenamefont {Fox}, \citenamefont {Follett}, \citenamefont
  {Fiksel}, \citenamefont {Li}, \citenamefont {Matteucci}, \citenamefont
  {Bhattacharjee},\ and\ \citenamefont {Germaschewski}}]{SchaefferPRL2019}%
  \BibitemOpen
  \bibfield  {author} {\bibinfo {author} {\bibfnamefont {D.~B.}\ \bibnamefont
  {Schaeffer}}, \bibinfo {author} {\bibfnamefont {W.}~\bibnamefont {Fox}},
  \bibinfo {author} {\bibfnamefont {R.~K.}\ \bibnamefont {Follett}}, \bibinfo
  {author} {\bibfnamefont {G.}~\bibnamefont {Fiksel}}, \bibinfo {author}
  {\bibfnamefont {C.~K.}\ \bibnamefont {Li}}, \bibinfo {author} {\bibfnamefont
  {J.}~\bibnamefont {Matteucci}}, \bibinfo {author} {\bibfnamefont
  {A.}~\bibnamefont {Bhattacharjee}},\ and\ \bibinfo {author} {\bibfnamefont
  {K.}~\bibnamefont {Germaschewski}},\ }\bibfield  {title} {\bibinfo {title}
  {Direct observations of particle dynamics in magnetized collisionless shock
  precursors in laser-produced plasmas},\ }\href
  {https://doi.org/10.1103/physrevlett.122.245001} {\bibfield  {journal}
  {\bibinfo  {journal} {Phys.\ Rev.\ Lett.}\ }\textbf {\bibinfo {volume}
  {122}},\ \bibinfo {pages} {245001} (\bibinfo {year} {2019})}\BibitemShut
  {NoStop}%
\bibitem [{\citenamefont {Tzeferacos}\ \emph {et~al.}(2018)\citenamefont
  {Tzeferacos}, \citenamefont {Rigby}, \citenamefont {Bott}, \citenamefont
  {Bell}, \citenamefont {Bingham}, \citenamefont {Casner}, \citenamefont
  {Cattaneo}, \citenamefont {Churazov}, \citenamefont {Emig}, \citenamefont
  {Fiuza}, \citenamefont {Forest}, \citenamefont {Foster}, \citenamefont
  {Graziani}, \citenamefont {Katz}, \citenamefont {Koenig}, \citenamefont {Li},
  \citenamefont {Meinecke}, \citenamefont {Petrasso}, \citenamefont {Park},
  \citenamefont {Remington}, \citenamefont {Ross}, \citenamefont {Ryu},
  \citenamefont {Ryutov}, \citenamefont {White}, \citenamefont {Reville},
  \citenamefont {Miniati}, \citenamefont {Schekochihin}, \citenamefont {Lamb},
  \citenamefont {Froula},\ and\ \citenamefont
  {Gregori}}]{TzeferacosNatComm2018}%
  \BibitemOpen
  \bibfield  {author} {\bibinfo {author} {\bibfnamefont {P.}~\bibnamefont
  {Tzeferacos}}, \bibinfo {author} {\bibfnamefont {A.}~\bibnamefont {Rigby}},
  \bibinfo {author} {\bibfnamefont {A.~F.~A.}\ \bibnamefont {Bott}}, \bibinfo
  {author} {\bibfnamefont {A.~R.}\ \bibnamefont {Bell}}, \bibinfo {author}
  {\bibfnamefont {R.}~\bibnamefont {Bingham}}, \bibinfo {author} {\bibfnamefont
  {A.}~\bibnamefont {Casner}}, \bibinfo {author} {\bibfnamefont
  {F.}~\bibnamefont {Cattaneo}}, \bibinfo {author} {\bibfnamefont {E.~M.}\
  \bibnamefont {Churazov}}, \bibinfo {author} {\bibfnamefont {J.}~\bibnamefont
  {Emig}}, \bibinfo {author} {\bibfnamefont {F.}~\bibnamefont {Fiuza}},
  \bibinfo {author} {\bibfnamefont {C.~B.}\ \bibnamefont {Forest}}, \bibinfo
  {author} {\bibfnamefont {J.}~\bibnamefont {Foster}}, \bibinfo {author}
  {\bibfnamefont {C.}~\bibnamefont {Graziani}}, \bibinfo {author}
  {\bibfnamefont {J.}~\bibnamefont {Katz}}, \bibinfo {author} {\bibfnamefont
  {M.}~\bibnamefont {Koenig}}, \bibinfo {author} {\bibfnamefont {C.-K.}\
  \bibnamefont {Li}}, \bibinfo {author} {\bibfnamefont {J.}~\bibnamefont
  {Meinecke}}, \bibinfo {author} {\bibfnamefont {R.}~\bibnamefont {Petrasso}},
  \bibinfo {author} {\bibfnamefont {H.-S.}\ \bibnamefont {Park}}, \bibinfo
  {author} {\bibfnamefont {B.~A.}\ \bibnamefont {Remington}}, \bibinfo {author}
  {\bibfnamefont {J.~S.}\ \bibnamefont {Ross}}, \bibinfo {author}
  {\bibfnamefont {D.}~\bibnamefont {Ryu}}, \bibinfo {author} {\bibfnamefont
  {D.}~\bibnamefont {Ryutov}}, \bibinfo {author} {\bibfnamefont {T.~G.}\
  \bibnamefont {White}}, \bibinfo {author} {\bibfnamefont {B.}~\bibnamefont
  {Reville}}, \bibinfo {author} {\bibfnamefont {F.}~\bibnamefont {Miniati}},
  \bibinfo {author} {\bibfnamefont {A.~A.}\ \bibnamefont {Schekochihin}},
  \bibinfo {author} {\bibfnamefont {D.~Q.}\ \bibnamefont {Lamb}}, \bibinfo
  {author} {\bibfnamefont {D.~H.}\ \bibnamefont {Froula}},\ and\ \bibinfo
  {author} {\bibfnamefont {G.}~\bibnamefont {Gregori}},\ }\bibfield  {title}
  {\bibinfo {title} {Laboratory evidence of dynamo amplification of magnetic
  fields in a turbulent plasma},\ }\bibfield  {journal} {\bibinfo  {journal}
  {Nature Comm.}\ }\textbf {\bibinfo {volume} {9}},
  (\bibinfo {year} {2018})\BibitemShut {NoStop}%
\bibitem [{\citenamefont {Johnson}\ \emph {et~al.}(2022)\citenamefont
  {Johnson}, \citenamefont {Malko}, \citenamefont {Fox}, \citenamefont
  {Schaeffer}, \citenamefont {Fiksel}, \citenamefont {Adrian}, \citenamefont
  {Sutcliffe},\ and\ \citenamefont {Birkel}}]{JohnsonRSI2022}%
  \BibitemOpen
  \bibfield  {author} {\bibinfo {author} {\bibfnamefont {C.~L.}\ \bibnamefont
  {Johnson}}, \bibinfo {author} {\bibfnamefont {S.}~\bibnamefont {Malko}},
  \bibinfo {author} {\bibfnamefont {W.}~\bibnamefont {Fox}}, \bibinfo {author}
  {\bibfnamefont {D.~B.}\ \bibnamefont {Schaeffer}}, \bibinfo {author}
  {\bibfnamefont {G.}~\bibnamefont {Fiksel}}, \bibinfo {author} {\bibfnamefont
  {P.~J.}\ \bibnamefont {Adrian}}, \bibinfo {author} {\bibfnamefont {G.~D.}\
  \bibnamefont {Sutcliffe}},\ and\ \bibinfo {author} {\bibfnamefont
  {A.}~\bibnamefont {Birkel}},\ }\bibfield  {title} {\bibinfo {title} {Proton
  deflectometry with \textit{in situ} x-ray reference for absolute measurement
  of electromagnetic fields in high-energy-density plasmas},\ }\href
  {https://doi.org/10.1063/5.0064263} {\bibfield  {journal} {\bibinfo
  {journal} {Rev.\ Sci.\ Inst.}\ }\textbf {\bibinfo {volume}
  {93}},\ \bibinfo {pages} {023502} (\bibinfo {year} {2022})}\BibitemShut
  {NoStop}%
\bibitem [{\citenamefont {Malko}\ \emph {et~al.}(2022)\citenamefont {Malko},
  \citenamefont {Johnson}, \citenamefont {Schaeffer}, \citenamefont {Fox},\
  and\ \citenamefont {Fiksel}}]{MalkoApplOptics2022}%
  \BibitemOpen
  \bibfield  {author} {\bibinfo {author} {\bibfnamefont {S.}~\bibnamefont
  {Malko}}, \bibinfo {author} {\bibfnamefont {C.}~\bibnamefont {Johnson}},
  \bibinfo {author} {\bibfnamefont {D.~B.}\ \bibnamefont {Schaeffer}}, \bibinfo
  {author} {\bibfnamefont {W.}~\bibnamefont {Fox}},\ and\ \bibinfo {author}
  {\bibfnamefont {G.}~\bibnamefont {Fiksel}},\ }\bibfield  {title} {\bibinfo
  {title} {Design of proton deflectometry with in situ x-ray fiducial for
  magnetized high-energy-density systems},\ }\href
  {https://doi.org/10.1364/ao.448294} {\bibfield  {journal} {\bibinfo
  {journal} {Appl. Optics}\ }\textbf {\bibinfo {volume} {61}},\ \bibinfo
  {pages} {C133} (\bibinfo {year} {2022})}\BibitemShut {NoStop}%
\bibitem [{\citenamefont {Kasim}\ \emph {et~al.}(2017)\citenamefont {Kasim},
  \citenamefont {Ceurvorst}, \citenamefont {Ratan}, \citenamefont {Sadler},
  \citenamefont {Chen}, \citenamefont {S\"{a}vert}, \citenamefont {Trines},
  \citenamefont {Bingham}, \citenamefont {Burrows}, \citenamefont {Kaluza},\
  and\ \citenamefont {Norreys}}]{KasimPRE2017}%
  \BibitemOpen
  \bibfield  {author} {\bibinfo {author} {\bibfnamefont {M.~F.}\ \bibnamefont
  {Kasim}}, \bibinfo {author} {\bibfnamefont {L.}~\bibnamefont {Ceurvorst}},
  \bibinfo {author} {\bibfnamefont {N.}~\bibnamefont {Ratan}}, \bibinfo
  {author} {\bibfnamefont {J.}~\bibnamefont {Sadler}}, \bibinfo {author}
  {\bibfnamefont {N.}~\bibnamefont {Chen}}, \bibinfo {author} {\bibfnamefont
  {A.}~\bibnamefont {S\"{a}vert}}, \bibinfo {author} {\bibfnamefont
  {R.}~\bibnamefont {Trines}}, \bibinfo {author} {\bibfnamefont
  {R.}~\bibnamefont {Bingham}}, \bibinfo {author} {\bibfnamefont {P.~N.}\
  \bibnamefont {Burrows}}, \bibinfo {author} {\bibfnamefont {M.~C.}\
  \bibnamefont {Kaluza}},\ and\ \bibinfo {author} {\bibfnamefont
  {P.}~\bibnamefont {Norreys}},\ }\bibfield  {title} {\bibinfo {title}
  {Quantitative shadowgraphy and proton radiography for large intensity
  modulations},\ }\href {https://doi.org/10.1103/physreve.95.023306} {\bibfield
   {journal} {\bibinfo  {journal} {Phys.\ Rev. E}\ }\textbf {\bibinfo
  {volume} {95}},\ \bibinfo {pages} {023306} (\bibinfo {year}
  {2017})}\BibitemShut {NoStop}%
\bibitem [{\citenamefont {Davies}\ \emph {et~al.}(2023)\citenamefont {Davies},
  \citenamefont {Heuer},\ and\ \citenamefont {Bott}}]{DaviesHEDP2023}%
  \BibitemOpen
  \bibfield  {author} {\bibinfo {author} {\bibfnamefont {J.}~\bibnamefont
  {Davies}}, \bibinfo {author} {\bibfnamefont {P.}~\bibnamefont {Heuer}},\ and\
  \bibinfo {author} {\bibfnamefont {A.}~\bibnamefont {Bott}},\ }\bibfield
  {title} {\bibinfo {title} {Quantitative proton radiography and shadowgraphy
  for arbitrary intensities},\ }\href
  {https://doi.org/10.1016/j.hedp.2023.101067} {\bibfield  {journal} {\bibinfo
  {journal} {High Energy Density Phys.}\ }\textbf {\bibinfo {volume} {49}},\
  \bibinfo {pages} {101067} (\bibinfo {year} {2023})}\BibitemShut {NoStop}%
\bibitem [{\citenamefont {Sulman}\ \emph {et~al.}(2011)\citenamefont {Sulman},
  \citenamefont {Williams},\ and\ \citenamefont {Russell}}]{SulmanANM2011}%
  \BibitemOpen
  \bibfield  {author} {\bibinfo {author} {\bibfnamefont {M.~M.}\ \bibnamefont
  {Sulman}}, \bibinfo {author} {\bibfnamefont {J.}~\bibnamefont {Williams}},\
  and\ \bibinfo {author} {\bibfnamefont {R.~D.}\ \bibnamefont {Russell}},\
  }\bibfield  {title} {\bibinfo {title} {An efficient approach for the
  numerical solution of the {Monge--Amp{\`e}re} equation},\ }\href
  {https://doi.org/10.1016/j.apnum.2010.10.006} {\bibfield  {journal} {\bibinfo
   {journal} {Appl. Num. Math.}\ }\textbf {\bibinfo {volume}
  {61}},\ \bibinfo {pages} {298} (\bibinfo {year} {2011})}\BibitemShut
  {NoStop}%
\bibitem [{\citenamefont {Chen}\ \emph {et~al.}(2017)\citenamefont {Chen},
  \citenamefont {Kasim}, \citenamefont {Ceurvorst}, \citenamefont {Ratan},
  \citenamefont {Sadler}, \citenamefont {Levy}, \citenamefont {Trines},
  \citenamefont {Bingham},\ and\ \citenamefont {Norreys}}]{ChenPRE2017}%
  \BibitemOpen
  \bibfield  {author} {\bibinfo {author} {\bibfnamefont {N.~F.~Y.}\
  \bibnamefont {Chen}}, \bibinfo {author} {\bibfnamefont {M.~F.}\ \bibnamefont
  {Kasim}}, \bibinfo {author} {\bibfnamefont {L.}~\bibnamefont {Ceurvorst}},
  \bibinfo {author} {\bibfnamefont {N.}~\bibnamefont {Ratan}}, \bibinfo
  {author} {\bibfnamefont {J.}~\bibnamefont {Sadler}}, \bibinfo {author}
  {\bibfnamefont {M.~C.}\ \bibnamefont {Levy}}, \bibinfo {author}
  {\bibfnamefont {R.}~\bibnamefont {Trines}}, \bibinfo {author} {\bibfnamefont
  {R.}~\bibnamefont {Bingham}},\ and\ \bibinfo {author} {\bibfnamefont
  {P.}~\bibnamefont {Norreys}},\ }\bibfield  {title} {\bibinfo {title} {Machine
  learning applied to proton radiography of high-energy-density plasmas},\
  }\href {https://doi.org/10.1103/physreve.95.043305} {\bibfield  {journal}
  {\bibinfo  {journal} {Phys.\ Rev. E}\ }\textbf {\bibinfo {volume} {95}},\
  \bibinfo {pages} {043305} (\bibinfo {year} {2017})}\BibitemShut {NoStop}%
\bibitem [{\citenamefont {Walsh}\ \emph {et~al.}(2020)\citenamefont {Walsh},
  \citenamefont {Chittenden}, \citenamefont {Hill},\ and\ \citenamefont
  {Ridgers}}]{WalshPoP2020}%
  \BibitemOpen
  \bibfield  {author} {\bibinfo {author} {\bibfnamefont {C.~A.}\ \bibnamefont
  {Walsh}}, \bibinfo {author} {\bibfnamefont {J.~P.}\ \bibnamefont
  {Chittenden}}, \bibinfo {author} {\bibfnamefont {D.~W.}\ \bibnamefont
  {Hill}},\ and\ \bibinfo {author} {\bibfnamefont {C.}~\bibnamefont
  {Ridgers}},\ }\bibfield  {title} {\bibinfo {title}
  {Extended-magnetohydrodynamics in under-dense plasmas},\ }\href
  {https://doi.org/10.1063/1.5124144} {\bibfield  {journal} {\bibinfo
  {journal} {Phys.\ Plasmas}\ }\textbf {\bibinfo {volume} {27}},\ \bibinfo
  {pages} {022103} (\bibinfo {year} {2020})}\BibitemShut {NoStop}%
\bibitem [{\citenamefont {Gomez}\ \emph {et~al.}(2020)\citenamefont {Gomez},
  \citenamefont {Slutz}, \citenamefont {Jennings}, \citenamefont {Ampleford},
  \citenamefont {Weis}, \citenamefont {Myers}, \citenamefont {Yager-Elorriaga},
  \citenamefont {Hahn}, \citenamefont {Hansen}, \citenamefont {Harding},
  \citenamefont {Harvey-Thompson}, \citenamefont {Lamppa}, \citenamefont
  {Mangan}, \citenamefont {Knapp}, \citenamefont {Awe}, \citenamefont
  {Chandler}, \citenamefont {Cooper}, \citenamefont {Fein}, \citenamefont
  {Geissel}, \citenamefont {Glinsky}, \citenamefont {Lewis}, \citenamefont
  {Ruiz}, \citenamefont {Ruiz}, \citenamefont {Savage}, \citenamefont {Schmit},
  \citenamefont {Smith}, \citenamefont {Styron}, \citenamefont {Porter},
  \citenamefont {Jones}, \citenamefont {Mattsson}, \citenamefont {Peterson},
  \citenamefont {Rochau},\ and\ \citenamefont {Sinars}}]{GomezPRL2020}%
  \BibitemOpen
  \bibfield  {author} {\bibinfo {author} {\bibfnamefont {M.~R.}\ \bibnamefont
  {Gomez}}, \bibinfo {author} {\bibfnamefont {S.~A.}\ \bibnamefont {Slutz}},
  \bibinfo {author} {\bibfnamefont {C.~A.}\ \bibnamefont {Jennings}}, \bibinfo
  {author} {\bibfnamefont {D.~J.}\ \bibnamefont {Ampleford}}, \bibinfo {author}
  {\bibfnamefont {M.~R.}\ \bibnamefont {Weis}}, \bibinfo {author}
  {\bibfnamefont {C.~E.}\ \bibnamefont {Myers}}, \bibinfo {author}
  {\bibfnamefont {D.~A.}\ \bibnamefont {Yager-Elorriaga}}, \bibinfo {author}
  {\bibfnamefont {K.~D.}\ \bibnamefont {Hahn}}, \bibinfo {author}
  {\bibfnamefont {S.~B.}\ \bibnamefont {Hansen}}, \bibinfo {author}
  {\bibfnamefont {E.~C.}\ \bibnamefont {Harding}}, \bibinfo {author}
  {\bibfnamefont {A.~J.}\ \bibnamefont {Harvey-Thompson}}, \bibinfo {author}
  {\bibfnamefont {D.~C.}\ \bibnamefont {Lamppa}}, \bibinfo {author}
  {\bibfnamefont {M.}~\bibnamefont {Mangan}}, \bibinfo {author} {\bibfnamefont
  {P.~F.}\ \bibnamefont {Knapp}}, \bibinfo {author} {\bibfnamefont {T.~J.}\
  \bibnamefont {Awe}}, \bibinfo {author} {\bibfnamefont {G.~A.}\ \bibnamefont
  {Chandler}}, \bibinfo {author} {\bibfnamefont {G.~W.}\ \bibnamefont
  {Cooper}}, \bibinfo {author} {\bibfnamefont {J.~R.}\ \bibnamefont {Fein}},
  \bibinfo {author} {\bibfnamefont {M.}~\bibnamefont {Geissel}}, \bibinfo
  {author} {\bibfnamefont {M.~E.}\ \bibnamefont {Glinsky}}, \bibinfo {author}
  {\bibfnamefont {W.~E.}\ \bibnamefont {Lewis}}, \bibinfo {author}
  {\bibfnamefont {C.~L.}\ \bibnamefont {Ruiz}}, \bibinfo {author}
  {\bibfnamefont {D.~E.}\ \bibnamefont {Ruiz}}, \bibinfo {author}
  {\bibfnamefont {M.~E.}\ \bibnamefont {Savage}}, \bibinfo {author}
  {\bibfnamefont {P.~F.}\ \bibnamefont {Schmit}}, \bibinfo {author}
  {\bibfnamefont {I.~C.}\ \bibnamefont {Smith}}, \bibinfo {author}
  {\bibfnamefont {J.~D.}\ \bibnamefont {Styron}}, \bibinfo {author}
  {\bibfnamefont {J.~L.}\ \bibnamefont {Porter}}, \bibinfo {author}
  {\bibfnamefont {B.}~\bibnamefont {Jones}}, \bibinfo {author} {\bibfnamefont
  {T.~R.}\ \bibnamefont {Mattsson}}, \bibinfo {author} {\bibfnamefont {K.~J.}\
  \bibnamefont {Peterson}}, \bibinfo {author} {\bibfnamefont {G.~A.}\
  \bibnamefont {Rochau}},\ and\ \bibinfo {author} {\bibfnamefont {D.~B.}\
  \bibnamefont {Sinars}},\ }\bibfield  {title} {\bibinfo {title} {Performance
  scaling in magnetized liner inertial fusion experiments},\ }\href
  {https://doi.org/10.1103/physrevlett.125.155002} {\bibfield  {journal}
  {\bibinfo  {journal} {Phys.\ Rev.\ Lett.}\ }\textbf {\bibinfo {volume}
  {125}},\ \bibinfo {pages} {155002} (\bibinfo {year} {2020})}\BibitemShut
  {NoStop}%
\bibitem [{\citenamefont {Campbell}\ \emph {et~al.}(2020)\citenamefont
  {Campbell}, \citenamefont {Walsh}, \citenamefont {Russell}, \citenamefont
  {Chittenden}, \citenamefont {Crilly}, \citenamefont {Fiksel}, \citenamefont
  {Nilson}, \citenamefont {Thomas}, \citenamefont {Krushelnick},\ and\
  \citenamefont {Willingale}}]{CampbellPRL2020}%
  \BibitemOpen
  \bibfield  {author} {\bibinfo {author} {\bibfnamefont {P.~T.}\ \bibnamefont
  {Campbell}}, \bibinfo {author} {\bibfnamefont {C.~A.}\ \bibnamefont {Walsh}},
  \bibinfo {author} {\bibfnamefont {B.~K.}\ \bibnamefont {Russell}}, \bibinfo
  {author} {\bibfnamefont {J.~P.}\ \bibnamefont {Chittenden}}, \bibinfo
  {author} {\bibfnamefont {A.}~\bibnamefont {Crilly}}, \bibinfo {author}
  {\bibfnamefont {G.}~\bibnamefont {Fiksel}}, \bibinfo {author} {\bibfnamefont
  {P.~M.}\ \bibnamefont {Nilson}}, \bibinfo {author} {\bibfnamefont {A.~G.~R.}\
  \bibnamefont {Thomas}}, \bibinfo {author} {\bibfnamefont {K.}~\bibnamefont
  {Krushelnick}},\ and\ \bibinfo {author} {\bibfnamefont {L.}~\bibnamefont
  {Willingale}},\ }\bibfield  {title} {\bibinfo {title} {Magnetic signatures of
  radiation-driven double ablation fronts},\ }\href
  {https://doi.org/10.1103/physrevlett.125.145001} {\bibfield  {journal}
  {\bibinfo  {journal} {Phys.\ Rev.\ Lett.}\ }\textbf {\bibinfo {volume}
  {125}},\ \bibinfo {pages} {145001} (\bibinfo {year} {2020})}\BibitemShut
  {NoStop}%
\bibitem [{PRO(2024)}]{PROBLEM_2024}%
  \BibitemOpen
  \href {https://github.com/flash-center/PROBLEM} {\bibinfo {title} {The
  {PROBLEM} reconstruction code, https://github.com/flash-center/problem}}
  (\bibinfo {year} {Accessed 6-2024})\BibitemShut {NoStop}%
\bibitem [{\citenamefont {Kasim}\ \emph {et~al.}(2019)\citenamefont {Kasim},
  \citenamefont {Bott}, \citenamefont {Tzeferacos}, \citenamefont {Lamb},
  \citenamefont {Gregori},\ and\ \citenamefont {Vinko}}]{KasimPRE2019}%
  \BibitemOpen
  \bibfield  {author} {\bibinfo {author} {\bibfnamefont {M.~F.}\ \bibnamefont
  {Kasim}}, \bibinfo {author} {\bibfnamefont {A.~F.~A.}\ \bibnamefont {Bott}},
  \bibinfo {author} {\bibfnamefont {P.}~\bibnamefont {Tzeferacos}}, \bibinfo
  {author} {\bibfnamefont {D.~Q.}\ \bibnamefont {Lamb}}, \bibinfo {author}
  {\bibfnamefont {G.}~\bibnamefont {Gregori}},\ and\ \bibinfo {author}
  {\bibfnamefont {S.~M.}\ \bibnamefont {Vinko}},\ }\bibfield  {title} {\bibinfo
  {title} {Retrieving fields from proton radiography without source profiles},\
  }\href {https://doi.org/10.1103/physreve.100.033208} {\bibfield  {journal}
  {\bibinfo  {journal} {Phys.\ Rev. E}\ }\textbf {\bibinfo {volume}
  {100}},\ \bibinfo {pages} {033208} (\bibinfo {year} {2019})}\BibitemShut
  {NoStop}%
\bibitem [{\citenamefont {Fox}\ \emph {et~al.}(2023)\citenamefont {Fox},
  \citenamefont {Schaeffer}, \citenamefont {Rosenberg}, \citenamefont {Fiksel},
  \citenamefont {Matteucci}, \citenamefont {Park}, \citenamefont {Bott},
  \citenamefont {Lezhnin}, \citenamefont {Bhattacharjee}, \citenamefont
  {Kalantar}, \citenamefont {Remington}, \citenamefont {Uzdensky},
  \citenamefont {Li}, \citenamefont {S\'eguin},\ and\ \citenamefont
  {Hu}}]{Fox2023}%
  \BibitemOpen
  \bibfield  {author} {\bibinfo {author} {\bibfnamefont {W.}~\bibnamefont
  {Fox}}, \bibinfo {author} {\bibfnamefont {D.}~\bibnamefont {Schaeffer}},
  \bibinfo {author} {\bibfnamefont {M.}~\bibnamefont {Rosenberg}}, \bibinfo
  {author} {\bibfnamefont {G.}~\bibnamefont {Fiksel}}, \bibinfo {author}
  {\bibfnamefont {J.}~\bibnamefont {Matteucci}}, \bibinfo {author}
  {\bibfnamefont {H.-S.}\ \bibnamefont {Park}}, \bibinfo {author}
  {\bibfnamefont {A.}~\bibnamefont {Bott}}, \bibinfo {author} {\bibfnamefont
  {K.}~\bibnamefont {Lezhnin}}, \bibinfo {author} {\bibfnamefont
  {A.}~\bibnamefont {Bhattacharjee}}, \bibinfo {author} {\bibfnamefont
  {D.}~\bibnamefont {Kalantar}}, \bibinfo {author} {\bibfnamefont
  {B.}~\bibnamefont {Remington}}, \bibinfo {author} {\bibfnamefont
  {D.}~\bibnamefont {Uzdensky}}, \bibinfo {author} {\bibfnamefont
  {C.}~\bibnamefont {Li}}, \bibinfo {author} {\bibfnamefont {F.}~\bibnamefont
  {S\'eguin}},\ and\ \bibinfo {author} {\bibfnamefont {S.}~\bibnamefont {Hu}},\
  }\bibfield  {title} {\bibinfo {title} {Fast magnetic reconnection in
  highly-extended current sheets at the {National Ignition Facility}},\
  }\href@noop {} {\bibfield  {journal} {\bibinfo  {journal} {submitted to Phys.
  Rev. Lett., available https://arxiv.org/abs/2003.06351}\ } (\bibinfo {year}
  {2023})}\BibitemShut {NoStop}%
\bibitem [{\citenamefont {Graziani}\ \emph {et~al.}(2017)\citenamefont
  {Graziani}, \citenamefont {Tzeferacos}, \citenamefont {Lamb},\ and\
  \citenamefont {Li}}]{GrazianiRSI2017}%
  \BibitemOpen
  \bibfield  {author} {\bibinfo {author} {\bibfnamefont {C.}~\bibnamefont
  {Graziani}}, \bibinfo {author} {\bibfnamefont {P.}~\bibnamefont
  {Tzeferacos}}, \bibinfo {author} {\bibfnamefont {D.~Q.}\ \bibnamefont
  {Lamb}},\ and\ \bibinfo {author} {\bibfnamefont {C.}~\bibnamefont {Li}},\
  }\bibfield  {title} {\bibinfo {title} {Inferring morphology and strength of
  magnetic fields from proton radiographs},\ }\href
  {https://doi.org/10.1063/1.5013029} {\bibfield  {journal} {\bibinfo
  {journal} {Rev.\ Sci.\ Inst.}\ }\textbf {\bibinfo {volume}
  {88}},\ \bibinfo {pages} {123507} (\bibinfo {year} {2017})}\BibitemShut
  {NoStop}%
\bibitem [{\citenamefont {Fox}\ \emph {et~al.}(2011)\citenamefont {Fox},
  \citenamefont {Bhattacharjee},\ and\ \citenamefont
  {Germaschewski}}]{FoxPRL2011}%
  \BibitemOpen
  \bibfield  {author} {\bibinfo {author} {\bibfnamefont {W.}~\bibnamefont
  {Fox}}, \bibinfo {author} {\bibfnamefont {A.}~\bibnamefont {Bhattacharjee}},\
  and\ \bibinfo {author} {\bibfnamefont {K.}~\bibnamefont {Germaschewski}},\
  }\bibfield  {title} {\bibinfo {title} {{Fast Magnetic Reconnection in
  Laser-Produced Plasma Bubbles}},\ }\href
  {https://doi.org/10.1103/physrevlett.106.215003} {\bibfield  {journal}
  {\bibinfo  {journal} {Phys.\ Rev.\ Lett.}\ }\textbf {\bibinfo {volume}
  {106}},\ \bibinfo {pages} {215003} (\bibinfo {year} {2011})}\BibitemShut
  {NoStop}%
\bibitem [{\citenamefont {Fox}\ \emph {et~al.}(2012)\citenamefont {Fox},
  \citenamefont {Bhattacharjee},\ and\ \citenamefont
  {Germaschewski}}]{FoxPoP2012b}%
  \BibitemOpen
  \bibfield  {author} {\bibinfo {author} {\bibfnamefont {W.}~\bibnamefont
  {Fox}}, \bibinfo {author} {\bibfnamefont {A.}~\bibnamefont {Bhattacharjee}},\
  and\ \bibinfo {author} {\bibfnamefont {K.}~\bibnamefont {Germaschewski}},\
  }\bibfield  {title} {\bibinfo {title} {{Magnetic reconnection in
  high-energy-density laser-produced plasmas}},\ }\href
  {https://doi.org/10.1063/1.3694119} {\bibfield  {journal} {\bibinfo
  {journal} {Phys.\ Plasmas}\ }\textbf {\bibinfo {volume} {19}},\ \bibinfo
  {pages} {056309} (\bibinfo {year} {2012})}\BibitemShut {NoStop}%
\bibitem [{\citenamefont {Lezhnin}\ \emph {et~al.}(2018)\citenamefont
  {Lezhnin}, \citenamefont {Fox}, \citenamefont {Matteucci}, \citenamefont
  {Schaeffer}, \citenamefont {Bhattacharjee}, \citenamefont {Rosenberg},\ and\
  \citenamefont {Germaschewski}}]{LezhninPoP2018}%
  \BibitemOpen
  \bibfield  {author} {\bibinfo {author} {\bibfnamefont {K.~V.}\ \bibnamefont
  {Lezhnin}}, \bibinfo {author} {\bibfnamefont {W.}~\bibnamefont {Fox}},
  \bibinfo {author} {\bibfnamefont {J.}~\bibnamefont {Matteucci}}, \bibinfo
  {author} {\bibfnamefont {D.~B.}\ \bibnamefont {Schaeffer}}, \bibinfo {author}
  {\bibfnamefont {A.}~\bibnamefont {Bhattacharjee}}, \bibinfo {author}
  {\bibfnamefont {M.~J.}\ \bibnamefont {Rosenberg}},\ and\ \bibinfo {author}
  {\bibfnamefont {K.}~\bibnamefont {Germaschewski}},\ }\bibfield  {title}
  {\bibinfo {title} {Regimes of magnetic reconnection in colliding
  laser-produced magnetized plasma bubbles},\ }\href
  {https://doi.org/10.1063/1.5044547} {\bibfield  {journal} {\bibinfo
  {journal} {Phys.\ Plasmas}\ }\textbf {\bibinfo {volume} {25}},\ \bibinfo
  {pages} {093105} (\bibinfo {year} {2018})}\BibitemShut {NoStop}%
\bibitem [{\citenamefont {Yamada}\ \emph {et~al.}(2010)\citenamefont {Yamada},
  \citenamefont {Kulsrud},\ and\ \citenamefont {Ji}}]{YamadaRMP2010}%
  \BibitemOpen
  \bibfield  {author} {\bibinfo {author} {\bibfnamefont {M.}~\bibnamefont
  {Yamada}}, \bibinfo {author} {\bibfnamefont {R.}~\bibnamefont {Kulsrud}},\
  and\ \bibinfo {author} {\bibfnamefont {H.}~\bibnamefont {Ji}},\ }\bibfield
  {title} {\bibinfo {title} {{Magnetic reconnection}},\ }\href
  {https://doi.org/10.1103/revmodphys.82.603} {\bibfield  {journal} {\bibinfo
  {journal} {Rev.\ Mod.\ Phys.}\ }\textbf {\bibinfo {volume} {82}},\
  \bibinfo {pages} {603} (\bibinfo {year} {2010})}\BibitemShut {NoStop}%
\bibitem [{\citenamefont {S{\'{e}}guin}\ \emph {et~al.}(2003)\citenamefont
  {S{\'{e}}guin}, \citenamefont {Frenje}, \citenamefont {Li}, \citenamefont
  {Hicks}, \citenamefont {Kurebayashi}, \citenamefont {Rygg}, \citenamefont
  {Schwartz}, \citenamefont {Petrasso}, \citenamefont {Roberts}, \citenamefont
  {Soures}, \citenamefont {Meyerhofer}, \citenamefont {Sangster}, \citenamefont
  {Knauer}, \citenamefont {Sorce}, \citenamefont {Glebov}, \citenamefont
  {Stoeckl}, \citenamefont {Phillips}, \citenamefont {Leeper}, \citenamefont
  {Fletcher},\ and\ \citenamefont {Padalino}}]{SeguinRSI2003}%
  \BibitemOpen
  \bibfield  {author} {\bibinfo {author} {\bibfnamefont {F.~H.}\ \bibnamefont
  {S{\'{e}}guin}}, \bibinfo {author} {\bibfnamefont {J.~A.}\ \bibnamefont
  {Frenje}}, \bibinfo {author} {\bibfnamefont {C.~K.}\ \bibnamefont {Li}},
  \bibinfo {author} {\bibfnamefont {D.~G.}\ \bibnamefont {Hicks}}, \bibinfo
  {author} {\bibfnamefont {S.}~\bibnamefont {Kurebayashi}}, \bibinfo {author}
  {\bibfnamefont {J.~R.}\ \bibnamefont {Rygg}}, \bibinfo {author}
  {\bibfnamefont {B.-E.}\ \bibnamefont {Schwartz}}, \bibinfo {author}
  {\bibfnamefont {R.~D.}\ \bibnamefont {Petrasso}}, \bibinfo {author}
  {\bibfnamefont {S.}~\bibnamefont {Roberts}}, \bibinfo {author} {\bibfnamefont
  {J.~M.}\ \bibnamefont {Soures}}, \bibinfo {author} {\bibfnamefont {D.~D.}\
  \bibnamefont {Meyerhofer}}, \bibinfo {author} {\bibfnamefont {T.~C.}\
  \bibnamefont {Sangster}}, \bibinfo {author} {\bibfnamefont {J.~P.}\
  \bibnamefont {Knauer}}, \bibinfo {author} {\bibfnamefont {C.}~\bibnamefont
  {Sorce}}, \bibinfo {author} {\bibfnamefont {V.~Y.}\ \bibnamefont {Glebov}},
  \bibinfo {author} {\bibfnamefont {C.}~\bibnamefont {Stoeckl}}, \bibinfo
  {author} {\bibfnamefont {T.~W.}\ \bibnamefont {Phillips}}, \bibinfo {author}
  {\bibfnamefont {R.~J.}\ \bibnamefont {Leeper}}, \bibinfo {author}
  {\bibfnamefont {K.}~\bibnamefont {Fletcher}},\ and\ \bibinfo {author}
  {\bibfnamefont {S.}~\bibnamefont {Padalino}},\ }\bibfield  {title} {\bibinfo
  {title} {Spectrometry of charged particles from inertial-confinement-fusion
  plasmas},\ }\href {https://doi.org/10.1063/1.1518141} {\bibfield  {journal}
  {\bibinfo  {journal} {Rev.\ Sci.\ Inst.}\ }\textbf {\bibinfo
  {volume} {74}},\ \bibinfo {pages} {975} (\bibinfo {year} {2003})}\BibitemShut
  {NoStop}%
\end{thebibliography}%

\appendix

\bigskip
\section{Reconstruction Code and Availability}

The algorithms above have been implemented in Matlab
and are available in a package called PRADICAMENT on Github at the URL \url{https://github.com/wrfox/PRADICAMENT}

The code works in the consistent unit scheme 
described near Eqs.~\ref{Eq_KB} and \ref{Eq_KE}.
That is, if $K_B$ is in T and spatial units are in m, then $b(x)$ will be in 
T-m.  If $K_B = 1$ is specified, then the
value returned from the reconstructions will be $\xi(x)$ rather than $b(x)$.
If $K_E$ is used rather than $K_B$ then the values returned will be line-integrated electric fields.
The spatial coordinates are that of the plasma plane, as described near 
Eq.~\ref{Eq_mapping}.

The main routines, as of v1.0 are:

\vspace{10pt}

\texttt{prad\_inv} --- proton-radiography inverse solver in 1-D,
using specified boundary conditions.   It takes as input data proton fluence $I$
and $I_0$, $K_B$, and a required
boundary condition pair $(x_1, b_1)$, and $(x_2, b_2)$.
$I_0$ provides only the overall shape, since it 
is first renormalized to achieve the specified
boundary conditions on $b(x)$, per Eq.~\ref{Eq_AvgI0}.

\texttt{prad\_inv\_I0} --- proton-radiography inverse solver in 1-D,
using a specified $I_0$.
It takes as input data $I$ and $I_0$ as a function of 
coordinate $x$, along with a specified $K_B$, and an
optional $(x_0, B_0)$ pair to initiate the integration.
It integrates Eqs.~\ref{Eq_Bderiv} coupled to Eq.~\ref{Eq_mapping},
using standard ODE solvers.  The
routine interpolates $I$ and $I_0$ between values at the specified
mesh points as needed.

\texttt{prad\_fwd} --- produces a forward model proton image $I_{fwd}(x)$
from a given magnetic field profile $b(x)$.
It launches a large number off synthetic protons which are sent through the mapping,
and binned to final positions.
Because it uses binning, it correctly produces the proton image even
in caustic regimes.
Required inputs are $I_0(x)$, $b(x)$, and $K_B$.

%

\end{document}